\documentclass{article}

\PassOptionsToPackage{numbers,sort&compress}{natbib}

\usepackage[preprint]{neurips_2026}

\usepackage[utf8]{inputenc}
\usepackage[T1]{fontenc}
\usepackage{hyperref}
\usepackage{url}
\usepackage{booktabs}
\usepackage{amsfonts}
\usepackage{nicefrac}
\usepackage{microtype}
\usepackage{amsmath}
\usepackage{placeins}
\usepackage{graphicx}
\usepackage{pifont}
\usepackage{fontawesome5}
\usepackage{enumitem}
\usepackage[dvipsnames,table]{xcolor}
\usepackage{makecell}
\usepackage{caption}      

\title{SWE-Explore: Benchmarking How Coding Agents Explore Repositories}
 
\author{
  Shaoqiu Zhang\textsuperscript{1}\thanks{Equal contribution},\enspace
  Yuhang Wang\textsuperscript{1}\footnotemark[1],\enspace
  Jialiang Liang\textsuperscript{2}\footnotemark[1],\enspace 
  Yuling Shi\textsuperscript{1},\enspace 
  Wenhao Zeng\textsuperscript{1},\enspace \\
  \textbf{Maoquan Wang}\textsuperscript{4},\enspace 
  \textbf{Shilin He}\textsuperscript{5},\enspace
  \textbf{Ningyuan Xu}\textsuperscript{4},\enspace
  \textbf{Siyu Ye}\textsuperscript{3},\enspace
  \textbf{Kai Cai}\textsuperscript{4},\enspace 
  \textbf{Xiaodong Gu}\textsuperscript{1}\thanks{Corresponding author: Xiaodong Gu, xiaodong.gu@sjtu.edu.cn}\\
  \textsuperscript{1}Shanghai Jiao Tong University\quad
  \textsuperscript{2}Xinjiang University \\
  \textsuperscript{3}University of Illinois at Urbana-Champaign\quad
  \textsuperscript{4}Independent Researcher\\
  \textsuperscript{5}The Chinese University of Hong Kong\\
}
\begin{document}

\maketitle
\begin{center}
\footnotesize
\faGithub\;\href{https://github.com/Qiushao-E/SWE-Explore-Bench}{\url{https://github.com/Qiushao-E/SWE-Explore-Bench}}\\[-0.3mm]
\raisebox{-0.15em}{\includegraphics[height=1.05em]{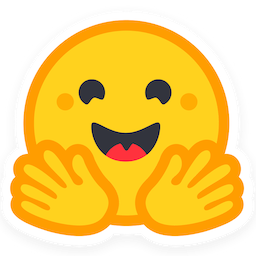}}\;\href{https://huggingface.co/datasets/SWE-Explore-Bench/SWE-Explore-Bench}{\url{https://huggingface.co/datasets/SWE-Explore-Bench/SWE-Explore-Bench}}
\end{center}
\begin{abstract}
Repository-level coding benchmarks such as SWE-bench have driven a rapid surge in the capabilities of coding agents. Yet they usually treat coding tasks as a holistic, binary prediction problem (e.g., resolved or unresolved), neglecting fine-grained agent capabilities such as repository understanding, context retrieval, code localization, and bug diagnosis. In this paper, we introduce \textbf{SWE-Explore}, a benchmark that isolates the evaluation of repository exploration, a critical capability of coding agents. Given a repository and an issue, SWE-Explore asks an explorer to return a ranked list of relevant code regions under a fixed line budget. SWE-Explore covers 848 issues across 10 programming languages and 203 open-source repositories. For each instance, we derive line-level ground truth from independent agent trajectories that successfully solved the same issue, distilling the specific code regions their solution paths actually consulted. We evaluate exploration along coverage, ranking, and context-efficiency dimensions, showing that these metrics strongly track downstream repair behavior. Across a broad set of retrieval methods, general coding agents, and specialized localizers, we find that agentic explorers form a clear tier above classical retrieval. While file-level localization is already strong for modern methods, line-level coverage and efficient ranking remain the key axes differentiating state-of-the-art explorers.
\end{abstract}

\section{Introduction}

Repository-level coding benchmarks, such as SWE-bench~\citep{jimenez2024swebench}, have driven a rapid surge in the capabilities of automated coding agents~\citep{chowdhury2024swebenchverified,zhang2025swebenchgoeslive,deng2025swebenchproaiagents}. The ecosystem around these benchmarks has expanded quickly: new evaluation distributions now cover multilingual repositories, multimodal software issues, and harder, long-horizon professional tasks~\citep{yang2025swesmith,zan2025multiswebenchmultilingualbenchmarkissue,yang2025swebench,badertdinov2025swerebenchautomatedpipelinetask}. In parallel, scalable training-oriented resources like SWE-smith~\citep{yang2025swesmith}, SWE-Gym~\citep{pan2025trainingsoftwareengineeringagents}, and SWE-Dev~\citep{wang2025swedev} are actively fueling agent development.Supported by these robust resources, frameworks such as SWE-agent~\citep{yang2024sweagent}, AutoCodeRover~\citep{zhang2024autocoderover}, Agentless~\citep{xia2024agentless}, OpenHands~\citep{wang2024openhands}, Claude Code~\citep{anthropic2025claudecode}, Mini-SWE-Agent~\citep{yang2024sweagent}, and AweAgent~\citep{chen2026beyondswecurrentcodeagent} have successfully turned repository-scale issue resolution into a practical, everyday testbed for software engineering agents.

\begin{figure}[t]
    \centering
     \includegraphics[
        width=\linewidth
    ]{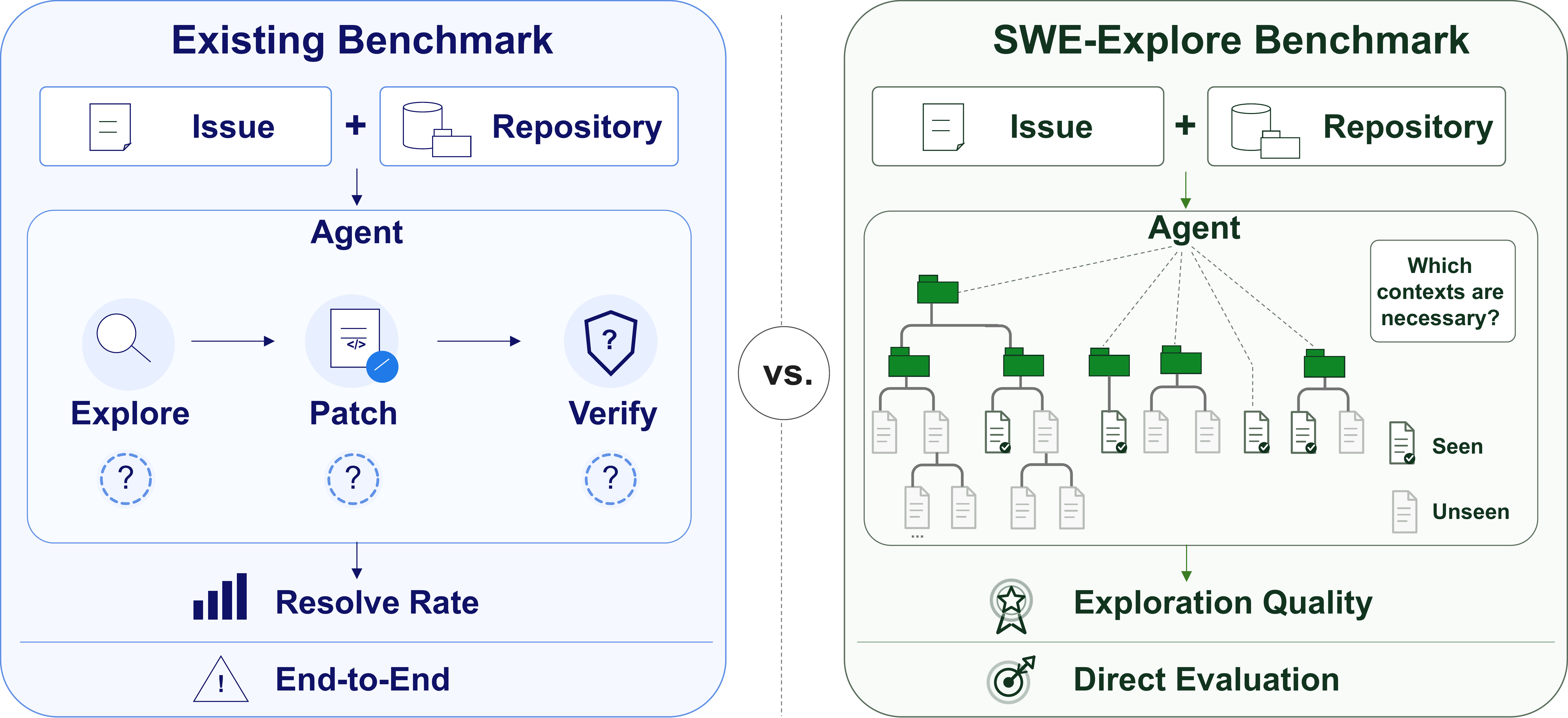}
    \caption{
    Motivation of SWE-Explore. A holistic metric of resolution rate conflates exploration, localization, and patch synthesis. SWE-Explore isolates repository exploration as a line-level evaluation target.}
    \label{fig:motivation}
    \vspace{-6mm}
\end{figure}

However, the widespread adoption of these benchmarks stems from a protocol that is both their strength and primary limitation: each repair attempt is reduced to a single pass/fail prediction, as shown in Figure~\ref{fig:motivation}. While this binary metric makes models directly comparable, it obscures the underlying mechanics of success. A holistic pass/fail score cannot reveal which specific step---reading the relevant code, localizing the bug, generating the patch, or validating the fix---actually succeeded or failed. Once we step back from this single prediction, two distinct failure modes emerge. An agent either fails to explore the relevant code for the fix, or it retrieves sufficient evidence but fails to synthesize a correct patch. While the latter is readily captured by existing executable benchmarks, the former is largely hidden. A real-world repository contains thousands of files. Determining which specific lines carries the evidence for a given issue is a daunting challenge, even for the agents that ultimately solve it. This decomposition is increasingly recognized by recent work on context management and bug localization~\citep{li2026contextbench,chen2025locagent,yu2025orcaloca,wang2026swepruner,zhu2026swecontextbench,suri2026codescout}.

Consequently, the capability of coding agents in repository exploration remains under-measured.
Despite recent efforts to study agentic localization and retrieval~\citep{li2026contextbench,zhang2025mulocbench,wang2026swepruner,zhu2026swecontextbench,chen2025locagent,yu2025orcaloca,suri2026codescout}, existing evaluations still lack a common, precise target for comparing classical retrievers, search agents, and long-context selectors. Measuring file or function level localization only indicates whether an agent reached the right general neighborhood, no metric reveals exactly \textit{which} lines of code were explored. Without visibility into line-level coverage, we cannot rigorously evaluate how well an agent explores a repository before it begins to write code.

In this paper, we introduce \textbf{SWE-Explore}, a benchmark that turns repository exploration into a comparable evaluation target. Given an issue and a repository, an explorer is asked to return a ranked list of code regions; we then score that list against ground truth derived from independent agent trajectories that successfully solved the same issue, asking how early it surfaces the evidence those trajectories actually relied on. The output format is deliberately simple: sparse retrievers, interactive agents, and long-context selectors are all compared as producers of the same ranked region list under a fixed line budget. This lets SWE-Explore evaluate exploration behavior without requiring the explorer to write or validate a patch.

To check that a higher exploration score actually leads to better repair, SWE-Explore is paired with a controlled downstream protocol: we feed each explorer's output---and only that output---as the available repository context to a fixed coding agent, and measure whether the resulting patch passes the original test suite. This protocol is not a replacement for the primary benchmark but a way to verify that what our exploration metrics measure is the same thing that drives downstream success. In this sense, the downstream protocol serves as an external validity check, while the benchmark itself remains a lightweight context-selection task.

In summary, our contributions are as follows:

\begin{itemize}[leftmargin=*,itemsep=2pt,topsep=0pt]

\item \textbf{A new evaluation target.} We isolate repository exploration from end-to-end repair and formalize it as a ranked, line-level context selection task, so that retrievers, search agents, and long-context selectors can be compared on the axis they are designed to improve.

\item \textbf{Trajectory-grounded supervision.} We propose a novel method to annotate ground truth lines from successful agent trajectories, with least requiring manual annotation.

\item \textbf{A studied metric set, validated against repair.} We systematically compare coverage, ranking, and budget-efficiency metrics, and use a controlled downstream protocol---where each explorer's output is the only context visible to a fixed coding agent---to show that the metrics we keep are predictive of repair success across a broad set of explorers.

\end{itemize}

\section{Related Work}
\subsection{Coding Benchmarks}

\providecommand{\cmkg}{\textcolor{green!55!black}{\ding{51}}}
\providecommand{\xmkr}{\textcolor{red!70!black}{\ding{55}}}

\begin{table}[t]
  
  \centering
  \caption{Comparison of SWE-Explore with existing repository-level coding and exploration benchmarks across six design dimensions covering ground-truth granularity, evaluation protocol, and ranked-region assessment.}
  \vspace{1mm}
  \footnotesize
  \renewcommand{\arraystretch}{1.18}
  \setlength{\tabcolsep}{4pt}
  \label{tab:compare}
  \begin{tabular*}{\columnwidth}{@{\extracolsep{\fill}} l cccccc @{}}
    \toprule
    \textbf{Benchmark}
      & \makecell{\textbf{Exec.}\\\textbf{Based}}
      & \makecell{\textbf{Multi-}\\\textbf{Lingual}}
      & \makecell{\textbf{Line-Level}\\\textbf{GT}}
      & \makecell{\textbf{Trajectory-}\\\textbf{Grounded GT}}
      & \makecell{\textbf{Joint Expl.}\\\textbf{+\,Repair Eval}}
      & \makecell{\textbf{Ranked}\\\textbf{Region Eval}} \\
    \midrule
    Loc-Bench~\citep{chen2025locagent}
      & \xmkr & \xmkr & \xmkr & \xmkr & \xmkr & \xmkr \\
    \midrule
    SWE-bench Verified~\citep{jimenez2024swebench,chowdhury2024swebenchverified}
      & \cmkg & \xmkr & \xmkr & \xmkr & \xmkr & \xmkr \\
    SWE-bench Multilingual~\citep{yang2025swesmith}
      & \cmkg & \cmkg & \xmkr & \xmkr & \xmkr & \xmkr \\
    SWE-bench-Pro~\citep{deng2025swebenchproaiagents}
      & \cmkg & \cmkg & \xmkr & \xmkr & \xmkr & \xmkr \\
    \midrule
    ContextBench~\citep{li2026contextbench}
      & \cmkg & \cmkg & \xmkr & \xmkr & \cmkg & \xmkr \\
    SWE-ContextBench~\citep{zhu2026swecontextbench}
      & \cmkg & \xmkr & \xmkr & \xmkr & \xmkr & \xmkr \\
    \midrule
    \textbf{SWE-Explore (Ours)}
      & \cmkg & \cmkg & \cmkg & \cmkg & \cmkg & \cmkg \\
    \bottomrule
  \end{tabular*}

  \vspace{-2mm}
\end{table}
Repository-level benchmarks have established executable issue resolution as the central evaluation target for coding agents. SWE-bench couples issue descriptions, repository snapshots, and harness-based verification \citep{jimenez2024swebench}, with Verified~\citep{chowdhury2024swebenchverified} and Live~\citep{zhang2025swebenchgoeslive} variants tightening quality and contamination control. Subsequent work broadens the scope along several axes: multilingual coverage (SWE-bench Multilingual\citep{yang2025swesmith}, Multi-SWE-bench\citep{zan2025multiswebenchmultilingualbenchmarkissue}), multi-turn and rebased settings (SWE-bench Multimodal\citep{yang2025swebench}, SWE-rebench\citep{badertdinov2025swerebenchautomatedpipelinetask}). A second line of work pushes evaluation toward intermediate behavior: ContextBench~\citep{li2026contextbench} introduces human-annotated gold contexts and scores retrieval over agent trajectories \citep{li2026contextbench}, while SWE-Pruner~\citep{wang2026swepruner}  and SWE-ContextBench~\citep{zhu2026swecontextbench} respectively benchmark context compression and experience reuse. Related efforts also examine adjacent repository-level abilities, including hierarchical debugging and correctness checking, programmer behavior patterns, multi-agent debate, experience-driven repair, and repository-level question answering~\citep{shi2024code,shi2025between,li2025swe,chen2025swe,peng2025swe}.

These benchmarks either target the full issue-to-patch pipeline or evaluate isolated facets of intermediate behavior; what is missing is a single benchmark in which trajectory-grounded, line-level exploration quality and its downstream effect on issue resolution can be measured jointly, as shown in Table~\ref{tab:compare}. This matters because exploration quality is not fully captured by either coarse context labels or final resolve rate: an explorer may reach the right file but miss the decisive span, or surface the right evidence too late in a ranked output. SWE-Explore is complementary rather than competing: it derives line-level supervision directly from successful agent trajectories, evaluates exploration as a ranked region-list task, and pairs the exploration score with a restricted-context executable validation on the same instances.

\begin{figure*}[t]
    \centering
    \includegraphics[
        width=0.98\textwidth
    ]{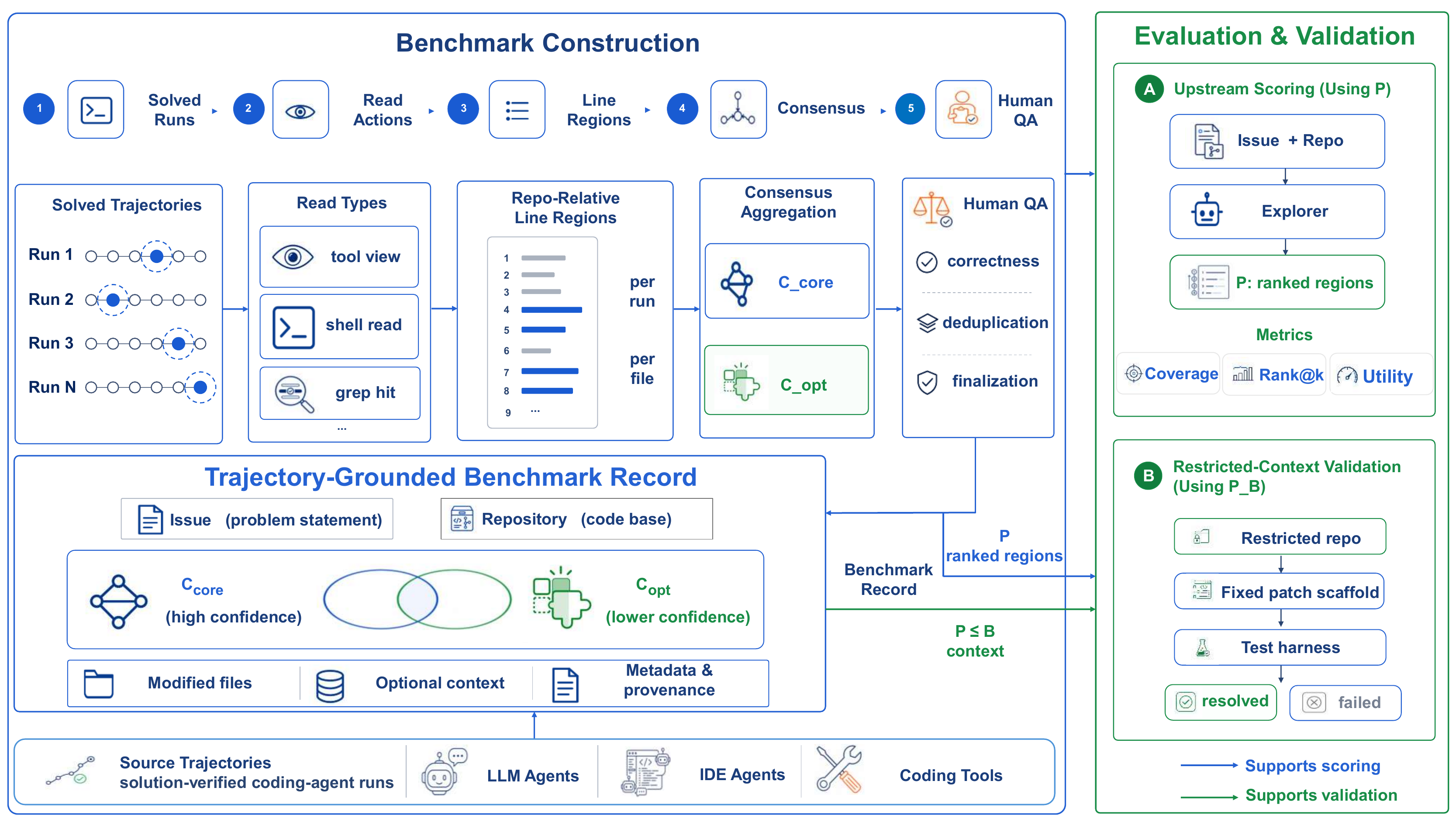}
    \caption{
    Overview of SWE-Explore. From solution-verified trajectories, SWE-Explore extracts read actions, aggregates them into core and optional context, and evaluates explorers using both upstream exploration metrics and downstream restricted-context validation.
    }
    \label{fig:framework}
    \vspace{-6mm}
\end{figure*}

\subsection{Explorer Methods}

Localization and explorer methods study how relevant code is found inside a repository. Classical retrieval and bug-localization work ranks code artifacts from natural-language reports, with TF--IDF~\citep{salton1988termweighting} and BM25~\citep{robertson2009bm25} as lightweight baselines and IR-based bug localization targeting likely faulty files~\citep{zhou2012buglocator}. Semantic code search and repository retrieval broaden this setting to natural-language-to-function search and cross-file code completion~\citep{husain2019codesearchnet,liu2024repobench}, while nDCG~\citep{jarvelin2002ndcg} provides a standard way to reward useful evidence appearing early in a ranked list. Long-context compression work further highlights that selecting, compressing, and preserving code context are themselves central design choices for code models and agents~\citep{shi2025longcodezip,wang2026context}. These evaluations supply useful methodology, but their targets are usually query--snippet relevance, next-line completion relevance, or bug-file relevance rather than the line regions consulted during successful issue resolution.

Recent LLM-based methods move from static retrieval toward interactive exploration. AutoCodeRover combines LLM reasoning, code search, and program analysis~\citep{zhang2024autocoderover}; LocAgent~\citep{chen2025locagent}, OrcaLoca~\citep{yu2025orcaloca}, and CoSIL~\citep{jiang2025issuelocalizationllmdriveniterative} evaluate localization over files, functions, ranked entities, or iterative code-graph search; and CodeScout studies pre-exploration for problem-statement improvement~\citep{suri2026codescout}. General-purpose coding agents further show that navigation, context management, tool use, and patch generation are tightly coupled in practical repair~\citep{yang2024sweagent,wang2024openhands,xia2024agentless,anthropic2025claudecode,chen2026beyondswecurrentcodeagent}. What remains less systematic is a common target for comparing lexical retrievers, dense retrievers, rerankers, and agentic explorers as ranked, line-level region producers. SWE-Explore fills this gap with trajectory-grounded line-level targets and a fixed-scaffold downstream protocol in which only the selected context varies.


\FloatBarrier

\section{SWE-Explore Benchmark}

\subsection{Task Formulation}
SWE-Explore formulates \emph{repository exploration} as a standalone functionality. Given an issue $q$ and a repository snapshot $\mathcal{R}$, SWE-Explore returns a ranked list of relevant code regions: 
\[
  f : (q, \mathcal{R}) \;\mapsto\; P = (r_1, r_2, \ldots, r_K),
\]
where \(P= (r_1, r_2, \ldots, r_K)\) is the \emph{ranked region list}. Each region $r_i = (p_i, s_i, e_i)$ consists of a file path $p_i$ and a line range $[s_i, e_i]$. SWE-Explore does not require a final patch, does not access the ground truth, and is not required to interact with the repository. As shown in Figure~\ref{fig:framework}, the benchmark scores $P$ against trajectory-grounded supervision derived per instance (\S\ref{sec:gt}) using the metric family in \S\ref{sec:metrics}. Independently, we use a restricted-context repair bridge (\S\ref{sec:bridge}) as a one-time methodological validation that these metrics track repair behavior; the bridge is not part of the standard evaluation loop, and a new explorer can be benchmarked using only the metrics above.

\subsection{Data Sources}
\label{sec:stats}
\begin{figure}[t]
  \centering
  \begin{minipage}[t]{0.42\linewidth}
    \vspace{0pt}
    \centering
    \includegraphics[width=\linewidth]{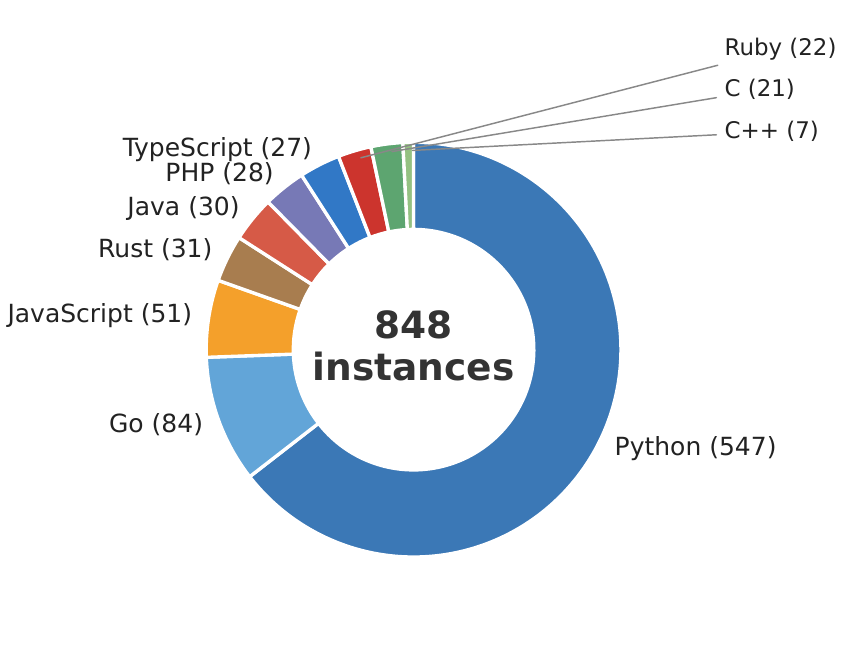}
    \captionof{figure}{Language distribution of the $848$ retained
    SWE-Explore instances across 10 different coding languages.}
    \label{fig:lang-dist}
  \end{minipage}\hfill
  \begin{minipage}[t]{0.55\linewidth}
    \vspace{0pt}
    \centering
    \captionof{table}{Per-instance averages of the ground-truth core context
    $|R_{\text{core}}|$ at the file, region, and line granularity.}
    \label{tab:bench-stats}
    \vspace{1mm}
    \footnotesize
\renewcommand{\arraystretch}{1.2}
\setlength{\tabcolsep}{6pt}
\begin{tabular}{@{} l l rr @{}}
  \toprule
  & & \textbf{Mean} & \textbf{Max} \\
  \midrule
  Issue Text & Length (Words) & 191.2 & 1,892 \\
  \midrule
  Ground-Truth & \# Files & 4.3 & 15 \\
  Context & \# Regions & 4.7 & 15 \\
   & \# Lines & 1,578 & 16,136 \\
  \midrule
  Provenance & \# Source Trajectories & 2.9 & 4 \\
   & \# Modified-by-Patch Files & 1.4 & 66 \\
  \midrule
  Codebase & \# Files (non-test) & 759 & 7,649 \\
   & \# Lines (non-test) & 179.6K & 1.4M \\
  \bottomrule
\end{tabular}

  \end{minipage}
  \vspace{-2mm}
\end{figure}

SWE-Explore is built on three public repository-level data sources: SWE-bench Verified~\citep{jimenez2024swebench,chowdhury2024swebenchverified}, SWE-bench-Pro~\citep{deng2025swebenchproaiagents}, and SWE-bench Multilingual~\citep{yang2025swesmith}. After the solution-verification filter described in \S\ref{sec:gt}, we retain $848$ instances spanning $10$ programming languages and $203$ open-source repositories. As Table~\ref{tab:bench-stats} and Figure~\ref{fig:lang-dist} summarizes, each instance carries on average $4.3$ ground-truth files, $4.7$ regions, and $1{,}578$ visible lines, embedded in repositories that average $759$ files and $180$K lines of non-test source code. Per-source breakdowns and the full benchmark distribution are deferred to Appendix~\ref{app:datasets}.

\subsection{Ground-Truth Annotation}
\label{sec:gt}

SWE-Explore keeps only instances for which we observe \emph{at least two successful issue-resolution trajectories} from strong LLMs such as GPT-5.4, Gemini-3-Pro, Sonnet-4.6, GLM-5.1, and Kimi-K2.6. Instances without two successful trajectories are excluded, because their trajectory-derived context would not support the cross-run agreement signal used below. After this filter, $848$ instances are retained across the three source datasets.

\paragraph{Trajectory Source.}
Directly annotating necessary context by hand is costly at our scale and difficult to make consistent: hundreds of repository-level issues span ten languages, and annotators may draw different boundaries around helpers, configuration, and tests. SWE-Explore instead derives supervision from \emph{solution-verified agent trajectories}: successful runs by strong coding agents such as GPT-5.4, Gemini-3-Pro, Sonnet-4.6, GLM-5.1, and Kimi-2.6 under the original harness. For each retained instance, we collect its successful trajectory set $T$ with $|T| \geq 2$.

We treat regions repeatedly surfaced across independent successful trajectories as a behavioral signal for core context: they are the code spans that different solution paths naturally explored while resolving the same issue. Given $T$, we first intersect read regions across trajectories to obtain conservative core candidates, then use an LLM-based refinement step to promote a small subset of model-specific optional reads when they are load-bearing for the issue. Finally, the authors manually audit every refined ground truth against the issue and trajectories, removing unsupported regions.

\paragraph{Extracting reads.} From each trajectory we collect all read actions that resolve to an explicit file--interval pair---editor-style \texttt{view} tool calls, command-line reads (\texttt{cat}/\texttt{head}/\texttt{tail}/\texttt{sed -n}), and \texttt{grep -n} line hits within $\mathcal{R}$---and normalize them into regions $(p, s, e)$. Actions we cannot unambiguously map to such a pair (e.g., free-form terminal interaction) are discarded rather than heuristically expanded, keeping the supervision record strictly grounded in observable reads.

\begin{figure}[t]
  \centering
  \includegraphics[width=\linewidth]{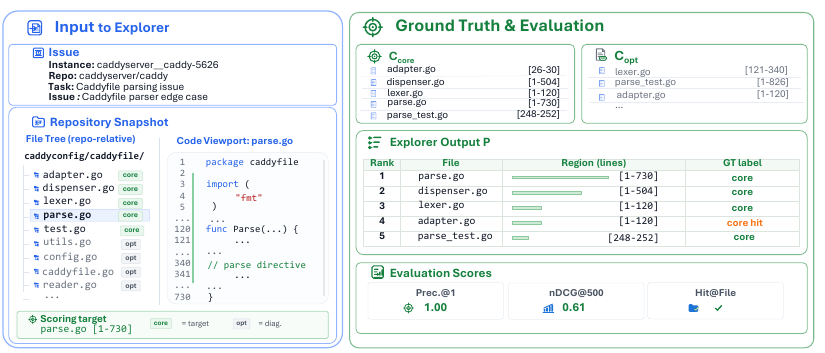}
  \caption{Example of a SWE-Explore instance. \textbf{Left:} an issue plus a repo snapshot with the highlighted core span. \textbf{Right:} trajectory-derived core regions $C_{\text{core}}$ (scoring target) and optional regions $C_{\text{opt}}$, an explorer's ranked prediction scored against $C_{\text{core}}$.}
  \label{fig:instance-example}
\end{figure}
\paragraph{Generating regions.}\label{sec:core-opt}
Let $R(\tau)$ be the set of $(p, s, e)$ regions extracted from trajectory $\tau$ as described above. We first compute a conservative intersection candidate $R_{\text{int}} = \bigcap_{\tau \in T} R(\tau)$ and collect model-specific optional reads outside this intersection as $R_{\text{opt}}^{(m)} = \bigl(\bigcup_{\tau \in T_m} R(\tau)\bigr) \setminus R_{\text{int}}$, where $T_m \subseteq T$ is the successful trajectories from model family $m$. Intersection and union are taken file-wise at the line level, so two overlapping reads of \texttt{parser.py:40--80} and \texttt{parser.py:60--100} contribute \texttt{parser.py:60--80} to $R_{\text{int}}$.

The final ground-truth core $R_{\text{core}}$ used in this paper is the \emph{refined} version of $R_{\text{int}}$: an LLM-based refinement step promotes a small subset of optional reads when they are load-bearing for the issue, and the authors then manually audit the resulting regions. Unless otherwise stated, all later uses of $R_{\text{core}}$ refer to this refined and audited ground truth, and $R_{\text{core}}$ is the only scoring target in the main experiments. Figure~\ref{fig:instance-example} shows an example instance of SWE-Explore Bench. Details and ablations comparing pure intersection, refined core, and union variants are deferred to Appendix~\ref{app:refine}.


\subsection{Metrics}
\label{sec:metrics}
 
With ground truth in hand, the next question is how to score an explorer's ranked region list against it. Let $L(r) \subseteq \{(p, \ell)\}$ be the set of $(p, \ell)$ pairs covered by region $r$, and for budget $B$ let $P_{\leq B}$ be the longest prefix of $P$ whose cumulative $|L(\cdot)|$ does not exceed $B$. Write $L(P) = \bigcup_i L(r_i)$ and $Y = L(R_{\text{core}})$.
 
\paragraph{Coverage and accuracy.}

\emph{Precision} and \emph{recall} are defined at the line level,
$\textsc{Prec} = |L(P) \cap Y| / |L(P)|$ and $\textsc{Rec} = |L(P) \cap Y| / |Y|$, with $\textsc{F1}$ their harmonic mean. We also report two coarser hit rates that capture the practically important event of overlapping the right code even when line spans are imprecise: a file-level
$\textsc{HitFile} = |\{p : \exists i,\, p_i = p\} \cap \textsc{Files}(Y)| / |\textsc{Files}(Y)|$,
and a region-level analogue
$\textsc{HitRegion} = |\{r \in R_{\text{core}} : \exists i,\, L(r_i) \cap L(r) \neq \emptyset\}| / |R_{\text{core}}|$,
which counts the fraction of core regions for which the explorer surfaced at least one overlapping prediction.
 
\paragraph{Ranking under budget.}
We adapt nDCG to a line-budget setting. Each predicted region $r_i$ is assigned a gain $g_i$ equal to the number of core lines it covers; regions are processed in their predicted order, and $\textsc{DCG@}B$ accumulates the discounted gain over the longest prefix whose cumulative line count stays within the budget $B$:
\[
  \textsc{DCG@}B \;=\; \sum_{i \in P_{\leq B}} \frac{g_i}{\log_2(i+2)}.
\]
$\textsc{nDCG@}B$ normalizes this against the best DCG attainable on the same instance under the same line budget; the ideal-ordering procedure is described in Appendix~\ref{app:idcg}. Using a line budget instead of a rank cutoff means a single verbose region that exhausts the budget without proportional gain is penalized just as heavily as omitting useful content. We additionally report \emph{first useful hit} (FUH), defined as $1 - i^\star / |P|$ where $i^\star$ is the smallest rank whose visible lines intersect the core target $Y$ (and $0$ if no rank in $P$ does); higher FUH means the explorer surfaced useful evidence earlier.

\paragraph{Efficiency and noise.}
\emph{Context efficiency} is the fraction of predicted visible lines that fall inside $L(R_{\text{core}}) \cup L(R_{\text{opt}}^{(m)})$, quantifying how much of the selected context is grounded evidence versus off-target. A complementary \emph{noise rate}, defined as the fraction of predicted regions overlapping neither $R_{\text{core}}$ nor $R_{\text{opt}}^{(m)}$, serves as a region-level diagnostic.

\paragraph{Validation by downstream repair.}\label{sec:bridge}
To check that the metrics above track downstream repair behavior, we construct a one-time \emph{restricted-context environment}: for a given explorer output $P$, we hide everything in the repository outside $\bigcup_i (p_i, [s_i, e_i])$, and ask a fixed coding agent to produce a patch that is then judged by the original SWE-bench harness. This is a one-time sanity check on the metrics and is not part of the standard evaluation procedure, and \S\ref{sec:downstream} uses it to quantify how well each metric predicts resolve rate; full implementation details---the sanitized container, the line-budget value $B$, the test-callback interface, and the patch scaffold---are given in Appendix~\ref{app:bridge}.

\section{Experiments}

\subsection{Setup}

\paragraph{Explorers.}
We evaluate explorers from four families. Two \emph{baselines} bound the dynamic range: Oracle returns $R_{\text{core}}$ directly, and Random returns uniformly sampled regions. \emph{Sparse retrievers} are represented by BM25~\citep{robertson2009bm25} and TF--IDF~\citep{salton1988termweighting}. As a \emph{lightweight dense retriever} we use a RAG pipeline instantiated with Potion, a static word-embedding retriever distilled from a sentence transformer. Finally, the \emph{agentic explorers} cover five general-purpose coding agents (Claude Code~\citep{anthropic2025claudecode}, Codex, OpenHands~\citep{wang2024openhands}, Mini-SWE-Agent~\citep{yang2024sweagent}, AweAgent~\citep{chen2026beyondswecurrentcodeagent}) and four published academic localization agents (AutoCodeRover~\citep{zhang2024autocoderover}, LocAgent~\citep{chen2025locagent}, OrcaLoca~\citep{yu2025orcaloca}, CoSIL~\citep{jiang2025issuelocalizationllmdriveniterative}); Oracle in particular plays a central role in validating our ground-truth construction (\S\ref{sec:downstream}).

\paragraph{Metrics.} Following the analysis in \S\ref{sec:metrics}, we report a combination of strongly predictive metrics and standard baselines. The primary metrics are Precision, nDCG@500, HitFile, and Context Efficiency, selected for their high correlation with downstream behavior and low mutual redundancy. We additionally report Recall, F1, and hit/noise region rates as conventional references, even though several of these exhibit weaker predictive power individually. Full metric definitions are in \S\ref{sec:metrics}.

\paragraph{Choice of $K$.}
On our refined ground truth, the per-instance number of core regions averages roughly $4.7$ after the LLM-refinement step (\S\ref{sec:gt}). We therefore fix $K{=}5$ for every explorer in this paper: each explorer is asked to return its five most relevant regions, which keeps the comparison both fair across systems and aligned with the size of the supervision target.

\begin{table}[t]
  \centering
  \begin{minipage}[t]{0.46\linewidth}
    \vspace{0pt}
    \centering
    \captionof{table}{Downstream resolve rate under the restricted-context
    validation environment (GPT-5.4 with Mini-SWE-Agent, $K{=}5$).}
    \label{tab:downstream}
    \vspace{1mm}
    \footnotesize
\renewcommand{\arraystretch}{1.15}
\begin{tabular}{@{} l c @{}}
  \toprule
  Explorer & Resolve Rate (\%) \\
  \midrule
  Oracle             & 59.7 \\
  Random             &  4.7 \\
  \midrule
  TF-IDF             & 26.0 \\
  RAG                & 23.3 \\
  BM25               & 12.7 \\
  \midrule
  CoSIL              & 59.3 \\
  Mini-SWE-Agent     & 50.0 \\
  Openhands          & 47.7 \\
  OrcaLoca           & 45.3 \\
  AutoCodeRover      & 44.7 \\
  LocAgent           & 44.7 \\
  AweAgent           & 41.3 \\
  \midrule
  Codex              & 50.3 \\
  Claude Code        & 48.0 \\
  \bottomrule
\end{tabular}

  \end{minipage}\hfill
  \begin{minipage}[t]{0.51\linewidth}
    \vspace{0pt}
    \centering
    \captionof{table}{Explorer-level correlation between each upstream
    exploration metric and downstream resolve rate, computed across all
    explorers in our pool. $\downarrow$ marks lower-is-better.}
    \label{tab:correlation}
    \vspace{1mm}
    \footnotesize
\renewcommand{\arraystretch}{1.12}
\setlength{\tabcolsep}{6pt}
\begin{tabular}{@{} l cc @{}}
  \toprule
  Metric & Pearson $r$ & Spearman $\rho$ \\
  \midrule
  CtxEff                 & $+0.950$ & $+0.739$ \\
  FUH                    & $+0.928$ & $+0.675$ \\
  Rec@100                & $+0.926$ & $+0.845$ \\
  HitFile                & $+0.925$ & $+0.695$ \\
  nDCG@500               & $+0.921$ & $+0.460$ \\
  nDCG@300               & $+0.920$ & $+0.458$ \\
  nDCG@100               & $+0.917$ & $+0.480$ \\
  HitReg                 & $+0.901$ & $+0.695$ \\
  Prec                   & $+0.890$ & $+0.671$ \\
  NoiseReg $\downarrow$  & $-0.812$ & $-0.562$ \\
  NoiseFile $\downarrow$ & $-0.808$ & $-0.590$ \\
  Rec@300                & $+0.769$ & $+0.796$ \\
  Rec@500                & $+0.710$ & $+0.796$ \\
  F1                     & $+0.673$ & $+0.810$ \\
  Rec$_\ell$             & $+0.617$ & $+0.796$ \\
  \bottomrule
\end{tabular}

  \end{minipage}
  \vspace{-2mm}
\end{table}

\begin{table*}[t]
  \centering
  \caption{Exploration quality at $K{=}5$ across different LLMs powering the same \textbf{Mini-SWE-Agent} scaffold. \textbf{Bold} marks the best result per column; \underline{underline} marks the second best.}
  \label{tab:llm-comparison}
  \vspace{1mm}
  \scriptsize
  \renewcommand{\arraystretch}{1.15}
  \begin{tabular*}{\textwidth}{@{\extracolsep{\fill}} l cccc c ccc cc @{}}
    \toprule
    & \multicolumn{5}{c}{\textit{Coverage \& Accuracy}} & \multicolumn{3}{c}{\textit{Ranking}} & \multicolumn{2}{c}{\textit{Efficiency \& Noise}} \\
    \cmidrule(lr){2-6} \cmidrule(lr){7-9} \cmidrule(lr){10-11}
    Model
      & HitReg & Prec & Rec$_\ell$ & F1 & HitFile
      & nDCG@500 & Rec@500 & FUH
      & CtxEff & NoiseReg$\downarrow$ \\
    \midrule
    GPT-5.4
      & \underline{0.516} & \textbf{0.542} & \underline{0.154} & \underline{0.194} & \textbf{0.655}
      & \underline{0.905} & \underline{0.154} & \underline{0.927}
      & \textbf{0.771} & \textbf{0.258} \\
    GPT-5.4-mini
      & \textbf{0.531} & 0.509 & \textbf{0.185} & \textbf{0.215} & \underline{0.649}
      & \textbf{0.924} & \textbf{0.183} & \textbf{0.956}
      & \underline{0.754} & \underline{0.265} \\
    Kimi-K2.6
      & 0.413 & 0.475 & 0.117 & 0.149 & 0.509
      & 0.739 & 0.115 & 0.759
      & 0.676 & 0.316 \\
    Sonnet-4.5
      & 0.428 & \underline{0.519} & {0.118} & 0.154 & 0.535
      & 0.779 & 0.116 & 0.802
      & 0.715 & 0.279 \\
    GLM-4.7
      & 0.289 & 0.414 & 0.122 & 0.148 & 0.343 & 0.557 & 0.105 & 0.572 & 0.536 & 0.465 \\
    Gemini-3-Pro
      & 0.268 & 0.420 & 0.052 & 0.079 & 0.369 & 0.605 & 0.052 & 0.620 & 0.540 & 0.467 \\
    \bottomrule
  \end{tabular*}
  \vspace{-4mm}
\end{table*}

\subsection{Downstream validation.}
\label{sec:downstream}

Before using upstream exploration metrics as the main evaluation target, we ask whether they track downstream repair behavior. On a shared $n{=}150$ subset of SWE-Explore Bench, each explorer provides its $K{=}5$ ranked regions; the resulting context is then given to a fixed Mini-SWE-Agent patcher backed by GPT-5.4 and Gemini-3-Pro, and we average the SWE-bench harness resolve rate over the two patchers. Table~\ref{tab:correlation} reports the explorer-level Pearson and Spearman correlations between each upstream metric and this downstream resolve rate.
 
\paragraph{Stable signals.}
The strongest metrics are not purely file-level or purely recall-based. Context Efficiency has the highest Pearson correlation ($r{=}0.950$), suggesting that useful context must be both relevant and compact. Rec@100 is the strongest rank-correlated signal ($\rho{=}0.845$), indicating that early coverage under a tight line budget is especially predictive of repair. HitFile, HitRegion, and FUH remain strong Pearson signals: they capture whether the explorer reaches the right file, overlaps the right evidence, and surfaces useful evidence early. Precision is also predictive, but it is most informative when interpreted together with coverage-oriented metrics.

\paragraph{Useful but incomplete signals.}
Rank-aware metrics such as nDCG@500 have very high Pearson correlation but weaker Spearman correlation, indicating that they separate broad quality tiers well but are less stable for ordering nearby explorers. Recall-style metrics show a budget effect: Rec@100 is strong in both views, while broader recall metrics and F1 are more rank-sensitive than scale-sensitive, reflecting their tendency to reward broad reading even when the selected context is not compact. Noise rates have the expected negative correlation, but serve better as diagnostics than as standalone success measures. Together, these results justify reporting a mixed metric set rather than a single score. In Table~\ref{tab:main-results} and Table~\ref{tab:llm-comparison}, we therefore emphasize HitRegion, HitFile, Precision, nDCG@500, FUH, and Context Efficiency, while retaining Recall, F1, Recall@500, and noise-region rate as complementary diagnostics.

\begin{table*}[t]
  \centering
  \caption{Exploration quality at $K{=}5$. \textbf{Bold} marks the best non-oracle result per column; \underline{underline} marks the second best. HitReg / HitFile are region-/file-level hit rates; Rec$_\ell$ is line-level recall. $\downarrow$ indicates lower is better; all others are higher-is-better. All agentic explorers are driven by GPT-5.4 as the underlying model.}
  
  \label{tab:main-results}
  \vspace{1mm}
  \scriptsize
  \renewcommand{\arraystretch}{1.10}
  \setlength{\tabcolsep}{4pt}
  \begin{tabular*}{\textwidth}{@{\extracolsep{\fill}} l ccccc ccc cc @{}}
    \toprule
    & \multicolumn{5}{c}{\textit{Coverage \& Accuracy}} & \multicolumn{3}{c}{\textit{Ranking}} & \multicolumn{2}{c}{\textit{Efficiency \& Noise}} \\
    \cmidrule(lr){2-6} \cmidrule(lr){7-9} \cmidrule(lr){10-11}
    Explorer
      & HitReg & Prec & Rec$_\ell$ & F1 & HitFile
      & nDCG@500 & Rec@500 & FUH
      & CtxEff & NoiseReg$\downarrow$ \\
    \midrule
    Oracle  & 0.915 & 1.000 & 0.953 & 0.964 & 0.923 & 0.858 & 0.576 & 1.000 & 1.000 & 0.000 \\
    Random  & 0.003 & 0.002 & 0.004 & 0.002 & 0.004 & 0.004 & 0.001 & 0.006 & 0.002 & 0.997 \\
    \midrule
    BM25    & 0.065 & 0.055 & 0.021 & 0.024 & 0.079 & 0.132 & 0.021 & 0.141 & 0.087 & 0.910 \\
    TF-IDF  & 0.121 & 0.117 & 0.049 & 0.054 & 0.140 & 0.223 & 0.049 & 0.240 & 0.190 & 0.821 \\
    Potion  & 0.069 & 0.055 & 0.025 & 0.026 & 0.088 & 0.136 & 0.025 & 0.146 & 0.100 & 0.897 \\
    \midrule
    OpenHands      & 0.514 & 0.489 & 0.179 & 0.209 & 0.645 & 0.867 & 0.177 & 0.895 & 0.737 & 0.245 \\
    Mini-SWE-Agent & 0.505 & 0.530 & 0.151 & 0.190 & 0.640 & 0.885 & 0.151 & 0.907 & 0.754 & 0.253 \\
    AweAgent       & \underline{0.534} & 0.577 & 0.140 & 0.182 & \textbf{0.682} & \textbf{0.954} & 0.140 & \underline{0.975} & \underline{0.829} & 0.191 \\
    AutoCodeRover  & 0.272 & \textbf{0.680} & \underline{0.233} & \underline{0.291} & 0.280 & 0.720 & 0.165 & 0.730 & 0.738 & \underline{0.034} \\
    LocAgent       & 0.472 & \underline{0.642} & 0.191 & 0.241 & 0.540 & \underline{0.950} & 0.173 & \textbf{0.977} & 0.799 & 0.195 \\
    OrcaLoca       & 0.126 & 0.295 & 0.033 & 0.049 & 0.129 & 0.311 & 0.030 & 0.313 & 0.317 & \textbf{0.003} \\
    CoSIL          & \textbf{0.544} & 0.581 & \textbf{0.788} & \textbf{0.602} & 0.544 & 0.824 & \textbf{0.412} & 0.920 & \textbf{0.898} & 0.471 \\
    \midrule
    Claude Code    & 0.531 & 0.598 & 0.154 & 0.202 & \underline{0.667} & 0.938 & 0.154 & 0.963 & \underline{0.829} & 0.186 \\
    Codex          & 0.516 & 0.523 & 0.194 & 0.223 & 0.649 & 0.901 & \underline{0.190} & 0.936 & 0.762 & 0.249 \\
    \bottomrule
  \end{tabular*}
  \vspace{-4mm}
\end{table*}

 
\subsection{Exploration Quality}
\label{sec:main-results}
 
Table~\ref{tab:llm-comparison} and Table~\ref{tab:main-results} report upstream exploration quality for all explorers and models at $K{=}5$. We highlight the following observations.

\paragraph{Agentic exploration is a clear step above non-agentic retrieval.}
Sparse retrievers (BM25, TF--IDF) and the lightweight dense retriever remain close to Random on most metrics, while every agentic explorer is substantially higher than them. This confirms that repository exploration is not well captured by one-shot lexical or embedding retrieval alone: multi-step interaction with the repository is already necessary to reach the metric range occupied by modern coding agents.
 
\paragraph{Low F1 is mostly a recall problem.}
Despite strong file-level hit rates and ranking scores, most non-oracle explorers still have low F1, and the limiting term is usually line-level recall rather than precision. The general-purpose coding agents all reach high HitFile and nDCG@500, but their Rec$_\ell$ remains only around $0.14$--$0.19$; AutoCodeRover is highly precise, yet also recall-limited. This suggests that broad enough repository exploration remains a central bottleneck for current code-agent-style explorers: they often find plausible files early, but miss many of the specific spans needed to cover the full ground-truth context.

\paragraph{LLM choice shifts the operating point, but not the bottleneck.}
Table~\ref{tab:llm-comparison} controls the scaffold by running the same Mini-SWE-Agent explorer with different LLMs. The GPT-family models form the strongest tier, but with slightly different profiles: GPT-5.4 is cleaner and more compact, while GPT-5.4-mini surfaces more core regions and ranks useful evidence earlier. Kimi-K2.6 and Sonnet-4.5 form a middle tier, and GLM-4.7 and Gemini-3-Pro lag mainly in coverage and ranking. The larger pattern is more important than the exact ordering: across all LLMs, file-level hits remain much stronger than line-level recall, so replacing the base model alone does not remove the exploration bottleneck. High-recall region discovery still appears to require better exploration mechanisms, not just a stronger patching model.

\paragraph{General coding agents behave surprisingly similarly.}
Claude Code, Codex, OpenHands, Mini-SWE-Agent, and AweAgent have closely matched profiles across coverage, ranking, and efficiency metrics. This is notable because they differ in implementation and harness complexity, yet their exploration outputs occupy nearly the same operating point: high file hit, high early ranking, compact context, and low line recall. The similarity suggests that a complex repair harness is not necessarily required to study the exploration subproblem; a simpler explorer interface can expose much of the same behavior.

\paragraph{Specialized localizers only help when they broaden search.}
The academic agents do not uniformly dominate general coding agents. AutoCodeRover is precise but conservative, OrcaLoca has very low noise but misses many relevant regions, and LocAgent resembles the general-agent profile rather than changing the recall frontier. CoSIL is the main exception: it achieves by far the highest non-oracle Rec$_\ell$ and F1, suggesting that its iterative code-graph search is an important component for high-recall exploration. In contrast, explorers that rely more heavily on shell-style navigation or narrow search actions may reach the right files while still under-covering line-level evidence.

\paragraph{Line-level evaluation adds information beyond file hits.}
HitFile remains a strong and useful signal, as shown by the downstream correlations in \S\ref{sec:downstream}; however, it does not distinguish whether an explorer actually surfaces the relevant spans inside those files. The contrast between high HitFile and much lower Rec$_\ell$ across most agentic explorers supports SWE-Explore's line-level design: file-level localization captures reaching the right neighborhood, while line-level metrics measure whether the evidence needed by successful trajectories is actually exposed.

\begin{figure}[t]
  \centering
  \includegraphics[width=0.95\linewidth]{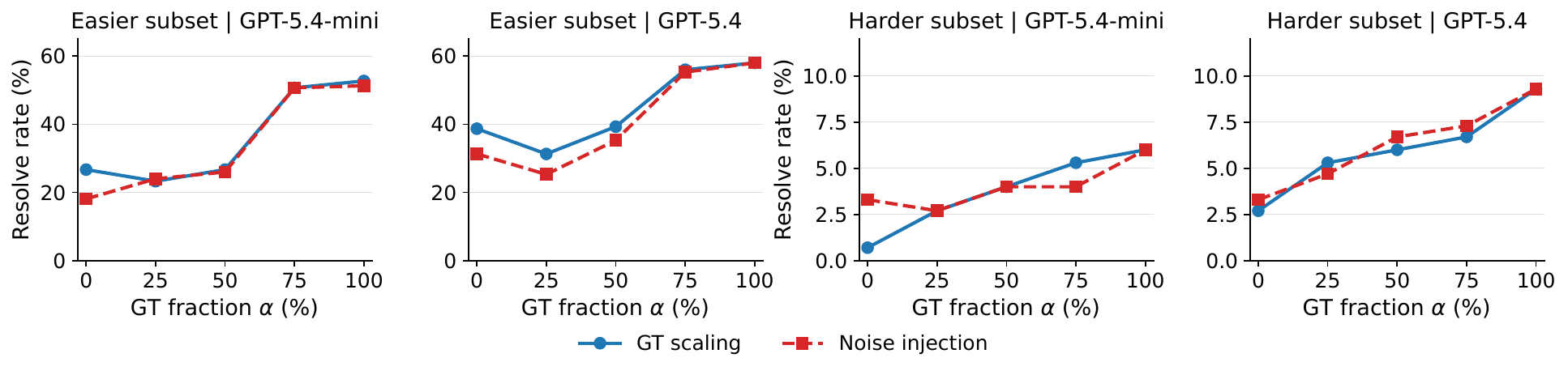}
  \vspace{-2mm}
  \caption{Resolve rate as the visible context degrades from the Oracle's full core set $R_{\text{core}}$ to either $\alpha\%$ of $R_{\text{core}}$ alone (\emph{GT scaling}, solid) or $\alpha\%$ of $R_{\text{core}}$ padded back to full size with random non-core regions (\emph{noise injection}, dashed).}
  \label{fig:degradation}
\end{figure}
\vspace{-3mm}
\subsection{Controlled Context Degradation}
\label{sec:degradation}
\vspace{-1mm}
The previous sections show that explorer behavior differs along coverage, precision, ranking, and efficiency, and that these differences affect downstream repair. Here we ask a more controlled robustness question: is a patcher more sensitive to \emph{missing} relevant context or to \emph{redundant} irrelevant context? In the restricted-context validation environment (\S\ref{sec:bridge}), we synthetically perturb Oracle context by exposing only an $\alpha\in\{0,25,50,75,100\}$ fraction of core regions (missing-context condition), or by filling the removed budget with randomly sampled non-core regions (redundant-context condition). We sweep $\alpha$ on two stratified $n{=}150$ subsets of SWE-Explore under both a weak patcher GPT-5.4-mini and a strong patcher GPT-5.4; Figure~\ref{fig:degradation} reports all four panels.

\vspace{-3mm}
\paragraph{Missing context is the dominant failure mode.}
On the easier subset, downstream resolve rate changes sharply only after enough core evidence is present: performance stays low through partial context, then jumps between $\alpha{=}50$ and $\alpha{=}75$. This threshold-like pattern suggests that patchers are not simply accumulating value smoothly from every additional region; instead, several pieces of core evidence must be present together before a correct fix becomes likely. Redundant context is less damaging once this threshold is crossed: the redundant-context curve closely tracks the missing-context curve for $\alpha\geq75$, indicating that modern patchers can tolerate extra irrelevant code when the essential evidence is already visible. This agrees with the metric analysis above: in the high-coverage regime, missing core evidence matters more than moderate precision loss, so recall-oriented improvements are more valuable than small filtering gains. Redundancy hurts most when core evidence is scarce, especially at $\alpha{=}0$, where random non-core code lowers resolve rate by $7$--$9$\,pp. The harder subset shows a much narrower range, suggesting that when the issue itself exceeds the patcher's capability, improving context alone has limited effect.

\vspace{-3mm}
\paragraph{The easier-subset dip suggests caution with empty-context baselines.}
Both easier-subset curves dip from $\alpha{=}0$ to $\alpha{=}25$, especially under the stronger patcher. A plausible explanation is memorization: with no repository context, the model may rely on an issue-only prior, while a small incomplete slice of $R_{\text{core}}$ pushes it to reconcile partial evidence. Because the dip disappears on the harder subset, we treat it as a caveat rather than the main effect: empty-context baselines on canonical repositories may be inflated and should be interpreted carefully.

\section{Conclusion}

We presented \textbf{SWE-Explore}, a benchmark for evaluating repository exploration independently from patch generation through ranked, line-level context selection. Using trajectory-derived supervision, SWE-Explore compares retrievers, search agents, and long-context selectors by the evidence they surface rather than only by final repair outcomes. Our experiments show that exploration metrics track downstream repair, that current agents are strong at finding relevant files but remain recall-limited at the line level, and that missing core evidence hurts more than moderate redundant context. We hope SWE-Explore provides a focused target for building explorers that read repositories more broadly and expose the spans repair agents actually need.

\bibliographystyle{plainnat}
\bibliography{references}

\begin{thebibliography}{37}
\providecommand{\natexlab}[1]{#1}
\providecommand{\url}[1]{\texttt{#1}}
\expandafter\ifx\csname urlstyle\endcsname\relax
  \providecommand{\doi}[1]{doi: #1}\else
  \providecommand{\doi}{doi: \begingroup \urlstyle{rm}\Url}\fi

\bibitem[{Anthropic}(2025)]{anthropic2025claudecode}
{Anthropic}.
\newblock Claude code: Ai-assisted coding in real-world codebases, 2025.
\newblock URL \url{https://claude.ai/code}.
\newblock Accessed: 2026-05.

\bibitem[Badertdinov et~al.(2025)Badertdinov, Golubev, Nekrashevich, Shevtsov, Karasik, Andriushchenko, Trofimova, Litvintseva, and Yangel]{badertdinov2025swerebenchautomatedpipelinetask}
Ibragim Badertdinov, Alexander Golubev, Maksim Nekrashevich, Anton Shevtsov, Simon Karasik, Andrei Andriushchenko, Maria Trofimova, Daria Litvintseva, and Boris Yangel.
\newblock Swe-rebench: An automated pipeline for task collection and decontaminated evaluation of software engineering agents, 2025.
\newblock URL \url{https://arxiv.org/abs/2505.20411}.

\bibitem[Chen et~al.(2026)Chen, Meng, Zhao, Li, Cheng, Song, Chen, Lin, Chen, Zhao, Song, Liu, Chen, Jia, and Wen]{chen2026beyondswecurrentcodeagent}
Guoxin Chen, Fanzhe Meng, Jiale Zhao, Minghao Li, Daixuan Cheng, Huatong Song, Jie Chen, Yuzhi Lin, Hui Chen, Xin Zhao, Ruihua Song, Chang Liu, Cheng Chen, Kai Jia, and Ji-Rong Wen.
\newblock Beyondswe: Can current code agent survive beyond single-repo bug fixing?, 2026.
\newblock URL \url{https://arxiv.org/abs/2603.03194}.

\bibitem[Chen et~al.(2025{\natexlab{a}})Chen, Lin, Shi, Lian, Gu, Yun, Chen, Cao, Liu, Xia, et~al.]{chen2025swe}
Silin Chen, Shaoxin Lin, Yuling Shi, Heng Lian, Xiaodong Gu, Longfei Yun, Dong Chen, Lin Cao, Jiyang Liu, Nu~Xia, et~al.
\newblock Swe-exp: Experience-driven software issue resolution.
\newblock \emph{arXiv preprint arXiv:2507.23361}, 2025{\natexlab{a}}.

\bibitem[Chen et~al.(2025{\natexlab{b}})Chen, Tang, Deng, Wu, Wu, Jiang, Prasanna, Cohan, and Wang]{chen2025locagent}
Zhaoling Chen, Xiangru Tang, Gangda Deng, Fang Wu, Jialong Wu, Zhiwei Jiang, Viktor~K. Prasanna, Arman Cohan, and Xingyao Wang.
\newblock {L}oc{A}gent: Graph-guided {LLM} agents for code localization.
\newblock \emph{CoRR}, abs/2503.09089, 2025{\natexlab{b}}.
\newblock \doi{10.48550/arXiv.2503.09089}.

\bibitem[Chowdhury et~al.(2024)Chowdhury, Aung, Shern, Jaffe, Sherburn, Starace, Mays, Dias, Aljubeh, Glaese, Jimenez, Yang, Ho, Patwardhan, Liu, and Madry]{chowdhury2024swebenchverified}
Neil Chowdhury, James Aung, Chan~Jun Shern, Oliver Jaffe, Dane Sherburn, Giulio Starace, Evan Mays, Rachel Dias, Marwan Aljubeh, Mia Glaese, Carlos~E. Jimenez, John Yang, Leyton Ho, Tejal Patwardhan, Kevin Liu, and Aleksander Madry.
\newblock Introducing {SWE}-bench verified, August 2024.
\newblock URL \url{https://openai.com/index/introducing-swe-bench-verified/}.
\newblock OpenAI milestone, updated February 24, 2025.

\bibitem[Deng et~al.(2025)Deng, Da, Pan, He, Ide, Garg, Lauffer, Park, Pasari, Rane, Sampath, Krishnan, Kundurthy, Hendryx, Wang, Bharadwaj, Holm, Aluri, Zhang, Jacobson, Liu, and Kenstler]{deng2025swebenchproaiagents}
Xiang Deng, Jeff Da, Edwin Pan, Yannis~Yiming He, Charles Ide, Kanak Garg, Niklas Lauffer, Andrew Park, Nitin Pasari, Chetan Rane, Karmini Sampath, Maya Krishnan, Srivatsa Kundurthy, Sean Hendryx, Zifan Wang, Vijay Bharadwaj, Jeff Holm, Raja Aluri, Chen Bo~Calvin Zhang, Noah Jacobson, Bing Liu, and Brad Kenstler.
\newblock Swe-bench pro: Can ai agents solve long-horizon software engineering tasks?, 2025.
\newblock URL \url{https://arxiv.org/abs/2509.16941}.

\bibitem[Husain et~al.(2019)Husain, Wu, Gazit, Allamanis, and Brockschmidt]{husain2019codesearchnet}
Hamel Husain, Ho-Hsiang Wu, Tiferet Gazit, Miltiadis Allamanis, and Marc Brockschmidt.
\newblock {CodeSearchNet} challenge: Evaluating the state of semantic code search, 2019.
\newblock URL \url{https://arxiv.org/abs/1909.09436}.

\bibitem[J{\"a}rvelin and Kek{\"a}l{\"a}inen(2002)]{jarvelin2002ndcg}
Kalervo J{\"a}rvelin and Jaana Kek{\"a}l{\"a}inen.
\newblock Cumulated gain-based evaluation of {IR} techniques.
\newblock \emph{ACM Transactions on Information Systems}, 20\penalty0 (4):\penalty0 422--446, 2002.
\newblock \doi{10.1145/582415.582418}.

\bibitem[Jiang et~al.(2025)Jiang, Ren, Yan, Jiang, Li, and Liu]{jiang2025issuelocalizationllmdriveniterative}
Zhonghao Jiang, Xiaoxue Ren, Meng Yan, Wei Jiang, Yong Li, and Zhongxin Liu.
\newblock Issue localization via llm-driven iterative code graph searching, 2025.
\newblock URL \url{https://arxiv.org/abs/2503.22424}.

\bibitem[Jimenez et~al.(2024)Jimenez, Yang, Wettig, Yao, Pei, Press, and Narasimhan]{jimenez2024swebench}
Carlos~E. Jimenez, John Yang, Alexander Wettig, Shunyu Yao, Kexin Pei, Ofir Press, and Karthik Narasimhan.
\newblock {SWE}-bench: Can language models resolve real-world {GitHub} issues?
\newblock In \emph{The Twelfth International Conference on Learning Representations}, 2024.
\newblock URL \url{https://openreview.net/forum?id=VTF8yNQM66}.

\bibitem[Li et~al.(2025)Li, Shi, Lin, Gu, Lian, Wang, Jia, Huang, and Wang]{li2025swe}
Han Li, Yuling Shi, Shaoxin Lin, Xiaodong Gu, Heng Lian, Xin Wang, Yantao Jia, Tao Huang, and Qianxiang Wang.
\newblock Swe-debate: Competitive multi-agent debate for software issue resolution.
\newblock \emph{arXiv preprint arXiv:2507.23348}, 2025.

\bibitem[Li et~al.(2026)Li, Zhu, Zhang, Feng, Wang, Pan, Barr, Sarro, Chu, and Ye]{li2026contextbench}
Han Li, Letian Zhu, Bohan Zhang, Rili Feng, Jiaming Wang, Yue Pan, Earl~T. Barr, Federica Sarro, Zhaoyang Chu, and He~Ye.
\newblock {C}ontext{B}ench: A benchmark for context retrieval in coding agents.
\newblock \emph{CoRR}, abs/2602.05892, 2026.
\newblock \doi{10.48550/arXiv.2602.05892}.

\bibitem[Liu et~al.(2024)Liu, Xu, and McAuley]{liu2024repobench}
Tianyang Liu, Canwen Xu, and Julian McAuley.
\newblock Repobench: Benchmarking repository-level code auto-completion systems.
\newblock In \emph{The Twelfth International Conference on Learning Representations}, 2024.
\newblock URL \url{https://openreview.net/forum?id=pPjZIOuQuF}.

\bibitem[Pan et~al.(2025)Pan, Wang, Neubig, Jaitly, Ji, Suhr, and Zhang]{pan2025trainingsoftwareengineeringagents}
Jiayi Pan, Xingyao Wang, Graham Neubig, Navdeep Jaitly, Heng Ji, Alane Suhr, and Yizhe Zhang.
\newblock Training software engineering agents and verifiers with swe-gym, 2025.
\newblock URL \url{https://arxiv.org/abs/2412.21139}.

\bibitem[Peng et~al.(2025)Peng, Shi, Wang, Zhang, Shen, and Gu]{peng2025swe}
Weihan Peng, Yuling Shi, Yuhang Wang, Xinyun Zhang, Beijun Shen, and Xiaodong Gu.
\newblock Swe-qa: Can language models answer repository-level code questions?
\newblock \emph{arXiv preprint arXiv:2509.14635}, 2025.

\bibitem[Robertson and Zaragoza(2009)]{robertson2009bm25}
Stephen Robertson and Hugo Zaragoza.
\newblock The probabilistic relevance framework: {BM25} and beyond.
\newblock \emph{Foundations and Trends in Information Retrieval}, 3\penalty0 (4):\penalty0 333--389, 2009.
\newblock \doi{10.1561/1500000019}.

\bibitem[Salton and Buckley(1988)]{salton1988termweighting}
Gerard Salton and Christopher Buckley.
\newblock Term-weighting approaches in automatic text retrieval.
\newblock \emph{Information Processing \& Management}, 24\penalty0 (5):\penalty0 513--523, 1988.
\newblock \doi{10.1016/0306-4573(88)90021-0}.

\bibitem[Shi et~al.(2024)Shi, Wang, Wan, Wang, and Gu]{shi2024code}
Yuling Shi, Songsong Wang, Chengcheng Wan, Min Wang, and Xiaodong Gu.
\newblock From code to correctness: Closing the last mile of code generation with hierarchical debugging.
\newblock \emph{arXiv preprint arXiv:2410.01215}, 2024.

\bibitem[Shi et~al.(2025{\natexlab{a}})Shi, Qian, Zhang, Shen, and Gu]{shi2025longcodezip}
Yuling Shi, Yichun Qian, Hongyu Zhang, Beijun Shen, and Xiaodong Gu.
\newblock Longcodezip: Compress long context for code language models.
\newblock In \emph{2025 40th IEEE/ACM International Conference on Automated Software Engineering (ASE)}, pages 141--153. IEEE, 2025{\natexlab{a}}.

\bibitem[Shi et~al.(2025{\natexlab{b}})Shi, Zhang, Wan, and Gu]{shi2025between}
Yuling Shi, Hongyu Zhang, Chengcheng Wan, and Xiaodong Gu.
\newblock Between lines of code: Unraveling the distinct patterns of machine and human programmers.
\newblock In \emph{2025 IEEE/ACM 47th International Conference on Software Engineering (ICSE)}, pages 1628--1639. IEEE, 2025{\natexlab{b}}.

\bibitem[Suri et~al.(2026)Suri, Li, Shojaie, Han, Hsu, Garg, Deshmukh, and Kumar]{suri2026codescout}
Manan Suri, Xiangci Li, Mehdi Shojaie, Songyang Han, Chao-Chun Hsu, Shweta Garg, Aniket~Anand Deshmukh, and Varun Kumar.
\newblock {C}ode{S}cout: Contextual problem statement enhancement for software agents.
\newblock \emph{CoRR}, abs/2603.05744, 2026.
\newblock \doi{10.48550/arXiv.2603.05744}.

\bibitem[Wang et~al.(2025)Wang, Hou, Wei, Tang, and Dong]{wang2025swedev}
Haoran Wang, Zhenyu Hou, Yao Wei, Jie Tang, and Yuxiao Dong.
\newblock {SWE}-dev: Building software engineering agents with training and inference scaling.
\newblock In \emph{Findings of the Association for Computational Linguistics: ACL 2025}, pages 3742--3761. Association for Computational Linguistics, 2025.
\newblock \doi{10.18653/v1/2025.findings-acl.193}.
\newblock URL \url{https://aclanthology.org/2025.findings-acl.193/}.

\bibitem[Wang et~al.(2024)Wang, Li, Song, Xu, Tang, Zhuge, Pan, Song, Li, Singh, Tran, Li, Ma, Zheng, Qian, Shao, Muennighoff, Zhang, Hui, Lin, Brennan, Peng, Ji, and Neubig]{wang2024openhands}
Xingyao Wang, Boxuan Li, Yufan Song, Frank~F. Xu, Xiangru Tang, Mingchen Zhuge, Jiayi Pan, Yueqi Song, Bowen Li, Jaskirat Singh, Hoang~H. Tran, Fuqiang Li, Ren Ma, Mingzhang Zheng, Bill Qian, Yanjun Shao, Niklas Muennighoff, Yizhe Zhang, Binyuan Hui, Junyang Lin, Robert Brennan, Hao Peng, Heng Ji, and Graham Neubig.
\newblock {OpenHands}: An open platform for {AI} software developers as generalist agents, 2024.
\newblock URL \url{https://arxiv.org/abs/2407.16741}.

\bibitem[Wang et~al.(2026{\natexlab{a}})Wang, Wang, Shi, Chen, Wang, Wang, Shen, Li, Gu, McAuley, et~al.]{wang2026context}
Yifei Wang, Ziteng Wang, Yuling Shi, Silin Chen, Xinrui Wang, Yueqi Wang, Beijun Shen, Linjing Li, Xiaodong Gu, Julian McAuley, et~al.
\newblock Context compression for llm agents: A survey of methods, failure modes, and evaluation.
\newblock 2026{\natexlab{a}}.

\bibitem[Wang et~al.(2026{\natexlab{b}})Wang, Shi, Yang, Zhang, He, Lian, Chen, Ye, Cai, and Gu]{wang2026swepruner}
Yuhang Wang, Yuling Shi, Mo~Yang, Rongrui Zhang, Shilin He, Heng Lian, Yuting Chen, Siyu Ye, Kai Cai, and Xiaodong Gu.
\newblock {SWE}-pruner: Self-adaptive context pruning for coding agents.
\newblock \emph{CoRR}, abs/2601.16746, 2026{\natexlab{b}}.
\newblock \doi{10.48550/arXiv.2601.16746}.

\bibitem[Xia et~al.(2024)Xia, Deng, Dunn, and Zhang]{xia2024agentless}
Chunqiu~Steven Xia, Yinlin Deng, Soren Dunn, and Lingming Zhang.
\newblock Agentless: Demystifying {LLM}-based software engineering agents.
\newblock \emph{CoRR}, abs/2407.01489, 2024.
\newblock \doi{10.48550/arXiv.2407.01489}.

\bibitem[Yang et~al.(2024)Yang, Jimenez, Wettig, Lieret, Yao, Narasimhan, and Press]{yang2024sweagent}
John Yang, Carlos~E. Jimenez, Alexander Wettig, Kilian Lieret, Shunyu Yao, Karthik~R. Narasimhan, and Ofir Press.
\newblock {SWE}-agent: Agent-computer interfaces enable automated software engineering.
\newblock In \emph{Advances in Neural Information Processing Systems}, 2024.
\newblock URL \url{https://openreview.net/forum?id=mXpq6ut8J3}.

\bibitem[Yang et~al.(2025{\natexlab{a}})Yang, Jimenez, Zhang, Lieret, Yang, Wu, Press, Muennighoff, Synnaeve, Narasimhan, Yang, Wang, and Press]{yang2025swebench}
John Yang, Carlos~E Jimenez, Alex~L Zhang, Kilian Lieret, Joyce Yang, Xindi Wu, Ori Press, Niklas Muennighoff, Gabriel Synnaeve, Karthik~R Narasimhan, Diyi Yang, Sida Wang, and Ofir Press.
\newblock {SWE}-bench multimodal: Do {AI} systems generalize to visual software domains?
\newblock In \emph{The Thirteenth International Conference on Learning Representations}, 2025{\natexlab{a}}.
\newblock URL \url{https://openreview.net/forum?id=riTiq3i21b}.

\bibitem[Yang et~al.(2025{\natexlab{b}})Yang, Lieret, Jimenez, Wettig, Khandpur, Zhang, Hui, Press, Schmidt, and Yang]{yang2025swesmith}
John Yang, Kilian Lieret, Carlos~E. Jimenez, Alexander Wettig, Kabir Khandpur, Yanzhe Zhang, Binyuan Hui, Ofir Press, Ludwig Schmidt, and Diyi Yang.
\newblock Swe-smith: Scaling data for software engineering agents, 2025{\natexlab{b}}.
\newblock URL \url{https://arxiv.org/abs/2504.21798}.

\bibitem[Yu et~al.(2025)Yu, Zhang, Zhao, Huang, Yao, Ding, and Zhao]{yu2025orcaloca}
Zhongming Yu, Hejia Zhang, Yujie Zhao, Hanxian Huang, Matrix Yao, Ke~Ding, and Jishen Zhao.
\newblock {O}rca{L}oca: An {LLM} agent framework for software issue localization.
\newblock In \emph{International Conference on Machine Learning}, 2025.
\newblock URL \url{https://openreview.net/forum?id=LyUfPOvM6I}.

\bibitem[Zan et~al.(2025)Zan, Huang, Liu, Chen, Zhang, Xin, Chen, Liu, Zhong, Li, Liu, Xiao, Chen, Zhang, Su, Liu, Long, Shen, and Xiang]{zan2025multiswebenchmultilingualbenchmarkissue}
Daoguang Zan, Zhirong Huang, Wei Liu, Hanwu Chen, Linhao Zhang, Shulin Xin, Lu~Chen, Qi~Liu, Xiaojian Zhong, Aoyan Li, Siyao Liu, Yongsheng Xiao, Liangqiang Chen, Yuyu Zhang, Jing Su, Tianyu Liu, Rui Long, Kai Shen, and Liang Xiang.
\newblock Multi-swe-bench: A multilingual benchmark for issue resolving, 2025.
\newblock URL \url{https://arxiv.org/abs/2504.02605}.

\bibitem[Zhang et~al.(2025{\natexlab{a}})Zhang, He, Zhang, Kang, Li, Xie, Wang, Wang, Huang, Fu, Nallipogu, Lin, Dang, Rajmohan, and Zhang]{zhang2025swebenchgoeslive}
Linghao Zhang, Shilin He, Chaoyun Zhang, Yu~Kang, Bowen Li, Chengxing Xie, Junhao Wang, Maoquan Wang, Yufan Huang, Shengyu Fu, Elsie Nallipogu, Qingwei Lin, Yingnong Dang, Saravan Rajmohan, and Dongmei Zhang.
\newblock Swe-bench goes live!, 2025{\natexlab{a}}.
\newblock URL \url{https://arxiv.org/abs/2505.23419}.

\bibitem[Zhang et~al.(2024)Zhang, Ruan, Fan, and Roychoudhury]{zhang2024autocoderover}
Yuntong Zhang, Haifeng Ruan, Zhiyu Fan, and Abhik Roychoudhury.
\newblock {A}uto{C}ode{R}over: Autonomous program improvement.
\newblock In \emph{Proceedings of the 33rd ACM SIGSOFT International Symposium on Software Testing and Analysis}, 2024.
\newblock \doi{10.1145/3650212.3680384}.

\bibitem[Zhang et~al.(2025{\natexlab{b}})Zhang, Wang, Yang, Pan, Tang, Li, Xing, Zhang, Li, and Zhang]{zhang2025mulocbench}
Zejun Zhang, Jian Wang, Qingyun Yang, Yifan Pan, Yi~Tang, Yi~Li, Zhenchang Xing, Tian Zhang, Xuandong Li, and Guoan Zhang.
\newblock {MULocBench}: A benchmark for localizing code and non-code issues in software projects, 2025{\natexlab{b}}.
\newblock URL \url{https://arxiv.org/abs/2509.25242}.

\bibitem[Zhou et~al.(2012)Zhou, Zhang, and Lo]{zhou2012buglocator}
Jian Zhou, Hongyu Zhang, and David Lo.
\newblock Where should the bugs be fixed? more accurate information retrieval-based bug localization based on bug reports.
\newblock In \emph{Proceedings of the 34th International Conference on Software Engineering}, pages 14--24. IEEE, 2012.
\newblock \doi{10.1109/ICSE.2012.6227210}.

\bibitem[Zhu et~al.(2026)Zhu, Hu, and Wu]{zhu2026swecontextbench}
Jared Zhu, Minhao Hu, and Junde Wu.
\newblock {SWE} context bench: A benchmark for context learning in coding.
\newblock \emph{CoRR}, abs/2602.08316, 2026.
\newblock \doi{10.48550/arXiv.2602.08316}.

\end{thebibliography}

\appendix
\appendix

\section{Dataset Details}
\label{app:datasets}

This appendix supplements \S\ref{sec:stats}. The main paper reports the aggregate
benchmark statistics; here we specify the retained instance set, record schema,
and repository-snapshot assumptions used by SWE-Explore.

\paragraph{Source benchmarks and filtering.}
SWE-Explore is constructed from three public repository-level sources:
SWE-bench Verified, SWE-bench-Pro, and SWE-bench Multilingual. We keep an
instance only when at least two trajectory resolves the
original task under the source benchmark's executable harness. This filtering
step ensures that the supervision target is extracted from successful repair
behavior rather than from failed exploration attempts. After filtering,
SWE-Explore contains 848 instances across 10 programming languages and 203
open-source repositories.

\paragraph{Benchmark composition.}
The retained set combines Python-centered verified issues, harder professional
software-engineering tasks, and multilingual issue-resolution tasks. This design
keeps the benchmark tied to executable repair while reducing dependence on a
single language ecosystem or repository family.

\paragraph{Language distribution.}
The language distribution is shown in Table~\ref{tab:language_distribution}.
Python remains the largest subset because SWE-bench Verified is Python-centric;
SWE-bench-Pro and SWE-bench Multilingual add most of the non-Python coverage.

\begin{table}[h]
\centering
\caption{Language distribution of SWE-Explore.}
\label{tab:language_distribution}
\begin{tabular}{lrr}
\toprule
Language & Instances & Percentage (\%) \\
\midrule
Python & 547 & 64.5 \\
Go & 84 & 9.9 \\
JavaScript & 51 & 6.0 \\
Rust & 31 & 3.7 \\
Java & 30 & 3.5 \\
PHP & 28 & 3.3 \\
TypeScript & 27 & 3.2 \\
Ruby & 22 & 2.6 \\
C & 21 & 2.5 \\
C++ & 7 & 0.8 \\
\midrule
Total & 848 & 100.0 \\
\bottomrule
\end{tabular}
\end{table}

\paragraph{Benchmark record schema.}
Each instance is stored as a structured record containing the issue, repository
metadata, trajectory provenance, and line-level supervision. The core fields are:
\begin{itemize}
    \item \texttt{instance\_id}: unique instance identifier;
    \item \texttt{repo}: source repository name;
    \item \texttt{source}: source benchmark;
    \item \texttt{problem\_statement}: natural-language issue description;
    \item \texttt{ground\_truth.read\_core\_regions}: line-level core regions used for scoring;
    \item \texttt{ground\_truth.read\_optional\_regions}: optional regions used for diagnostics and context-efficiency computation;
    \item \texttt{provenance}: successful source trajectories and extraction metadata.
\end{itemize}
All file paths are stored as repository-relative paths. Line intervals are
1-indexed and closed. Before scoring, paths are canonicalized so that equivalent
spellings such as \texttt{./src/foo.py} and \texttt{src/foo.py} map to the same
repository-relative file.

\paragraph{Repository snapshots.}
Each instance is evaluated against a fixed repository snapshot inherited from
its source benchmark. The same snapshot is used to resolve line intervals, score
explorer predictions, and run restricted-context downstream validation.
Generated files, temporary files, external dependency paths, and files outside
the repository checkout are not treated as valid repository files.

\section{Ground-Truth Construction and Refinement Details}
\label{app:refine}
\label{app:gt_construction}

This appendix supplements \S\ref{sec:gt}. The main paper describes the
trajectory-grounded construction at a high level; here we record the extraction,
normalization, refinement, and audit rules used to produce the final line-level
targets.

\paragraph{Read extraction.}
We extract observable file-reading behavior and convert it into
repository-relative line regions. We parse three types of read signals:
\begin{itemize}
    \item \textbf{Editor views}: tool calls with an explicit file path and visible line range;
    \item \textbf{Command-line reads}: commands such as \texttt{cat}, \texttt{head}, \texttt{tail}, and \texttt{sed -n} when the target file can be resolved;
    \item \textbf{Search hits}: \texttt{grep -n} outputs that can be mapped to repository files and line numbers.
\end{itemize}
Signals that cannot be mapped to a unique file--interval pair are discarded.
This rule keeps the supervision tied to observable repository reads and avoids
expanding ambiguous terminal output into unsupported line regions.

\paragraph{Path normalization.}
Absolute paths are accepted only when they point inside the repository checkout.
Relative paths are normalized by removing redundant \texttt{./} segments,
resolving \texttt{..} segments whenever possible, and matching the result
against the repository file index. Reads mapped to multiple candidate files are
discarded. Reads mapped outside the repository are also discarded.

\paragraph{Line-interval normalization.}
Each read is converted into a tuple $(p,s,e)$, where $p$ is the
repository-relative path and $[s,e]$ is a 1-indexed closed interval. Whole-file
reads are expanded using the file's line count at the evaluated checkout.
Out-of-range intervals are clipped to valid file boundaries. Empty intervals are
removed. Adjacent or overlapping intervals from the same trajectory and file are
merged before cross-trajectory aggregation.

\paragraph{Core and optional context.}
Let $\mathcal{T}$ be the successful trajectory set for an instance and let
$R(\tau)$ denote the merged line regions read by trajectory $\tau$. The raw core
context is the file-wise, line-level intersection across successful trajectories:
\[
R^{\mathrm{raw}}_{\mathrm{core}}
=
\bigcap_{\tau \in \mathcal{T}} R(\tau).
\]
The raw optional context is the portion of successful reads outside this
intersection:
\[
R^{\mathrm{raw}}_{\mathrm{opt}}
=
\left(\bigcup_{\tau \in \mathcal{T}} R(\tau)\right)
\setminus R^{\mathrm{raw}}_{\mathrm{core}}.
\]
The union has high recall but includes exploratory detours, redundant file
openings, and model-specific context that may not be necessary for repair. The
intersection is more conservative: it keeps only evidence consulted across
successful solution paths. SWE-Explore therefore uses the refined core as the
main scoring target and keeps optional context for diagnostics and
context-efficiency computation.

\paragraph{LLM-assisted refinement.}
Pure intersection can under-cover cases where different successful agents use
different but equivalent evidence. We therefore consider optional regions that
are repeatedly visited, directly adjacent to core evidence, or close to modified
regions. For each candidate, the refinement model receives the issue statement,
the candidate region, nearby code context, and a compact summary of successful
trajectories. The output schema contains:
\begin{itemize}
    \item a binary decision on whether the candidate is load-bearing;
    \item a short rationale grounded in the issue and trajectory evidence;
    \item the precise line interval to promote into the refined core.
\end{itemize}
Candidates without a precise promoted interval are rejected.

\paragraph{Human audit.}
Every promoted region is manually audited against the issue, source
trajectories, and final patch. The audit checks whether:
\begin{itemize}
    \item the region exists in the repository at the evaluated checkout;
    \item the region is relevant to the issue rather than merely adjacent or frequently opened;
    \item including it rewards evidence that a successful solution plausibly relied on.
\end{itemize}
Regions that fail any check are removed. The audit keeps the target conservative
while recovering load-bearing context that pure intersection may miss.

\paragraph{Context variants.}
For analysis, we maintain three target variants: pure intersection, refined core,
and full union. The pure-intersection target is maximally conservative. The
full-union target is high-recall but noisy. The refined-core target balances
these two extremes and is the default target used in the main experiments.

\section{Metric Definitions and Ideal-Order Implementation}
\label{app:metrics}

This appendix supplements \S\ref{sec:metrics}. The main paper defines the metric
family used in the experiments; here we give the evaluator-level definitions,
including duplicate handling, budgeted prefixes, and the ideal-order computation
used by nDCG.

\paragraph{Line universe.}
All metrics are computed over repository-relative line identifiers $(p,\ell)$,
where $p$ is a normalized path and $\ell$ is a 1-indexed line number. A predicted
region contributes all visible lines in its clipped interval. Duplicate
predicted lines are counted once for set-based precision and recall, while the
original region order is preserved for rank-aware metrics.

Let $L(r)$ denote the set of line identifiers covered by region $r$, let
$L(P)=\bigcup_i L(r_i)$ denote the union of predicted lines, and let
$Y=L(R_{\mathrm{core}})$ denote the line-level core target.

\paragraph{Coverage metrics.}
Line-level precision and recall are defined as:
\[
\mathrm{Prec}
=
\frac{|L(P)\cap Y|}{|L(P)|},
\qquad
\mathrm{Rec}_{\ell}
=
\frac{|L(P)\cap Y|}{|Y|}.
\]
F1 is the harmonic mean of precision and recall:
\[
\mathrm{F1}
=
\frac{2\cdot \mathrm{Prec}\cdot \mathrm{Rec}_{\ell}}
{\mathrm{Prec}+\mathrm{Rec}_{\ell}}.
\]

\paragraph{Hit rates.}
We also report two coarser hit rates. \textsc{HitFile} measures whether the
prediction reaches ground-truth files:
\[
\mathrm{HitFile}
=
\frac{
|\{p:\exists i, p_i=p\}\cap \mathrm{Files}(Y)|
}{
|\mathrm{Files}(Y)|
}.
\]
\textsc{HitRegion} measures whether each ground-truth region is overlapped by
at least one predicted region:
\[
\mathrm{HitRegion}
=
\frac{
|\{r\in R_{\mathrm{core}}:\exists i,\ L(r_i)\cap L(r)\neq \emptyset\}|
}{
|R_{\mathrm{core}}|
}.
\]

\paragraph{Budgeted prefixes.}
For a line budget $B$, $P_{\le B}$ is the longest prediction prefix whose
cumulative visible lines do not exceed $B$. This prefix definition penalizes
explorers that place very large regions early: a verbose early region can
exhaust the budget before more useful evidence appears. The main experiments use
$B=500$ for the primary rank-aware score and additionally compute
$B\in\{100,300,500\}$ in the released evaluator.

\paragraph{nDCG.}
For nDCG, each predicted region receives gain equal to the number of newly
covered core lines. Regions are processed in predicted order, and the discounted
gain is:
\[
\mathrm{DCG}@B
=
\sum_{i\in P_{\le B}}
\frac{g_i}{\log_2(i+2)}.
\]
The normalized score is:
\[
\mathrm{nDCG}@B
=
\frac{\mathrm{DCG}@B}{\mathrm{IDCG}@B}.
\]

\paragraph{Ideal-ordering for nDCG.}
\label{app:idcg}
The ideal DCG is computed under the same line budget as the explorer output. We
construct the ideal order greedily: at each step, the evaluator selects the
remaining ground-truth region with the largest marginal uncovered-line gain,
subject to the remaining line budget. Ties are broken by shorter region length
and then by repository path. This makes the normalization instance-specific and
budget-matched: an explorer is compared against the best achievable ordering of
the same target evidence, not against an unconstrained full-context oracle.

\paragraph{First useful hit.}
First Useful Hit (FUH) measures how early the explorer first surfaces any core
evidence. Let $i^\star$ be the first predicted rank whose visible lines intersect
$Y$. We define:
\[
\mathrm{FUH}
=
\begin{cases}
1 - i^\star/|P|, & \text{when such } i^\star \text{ exists},\\
0, & \text{otherwise}.
\end{cases}
\]
Higher values indicate that useful evidence appears earlier in the ranked list.

\paragraph{Efficiency and noise.}
Context efficiency is the fraction of predicted visible lines that fall inside
either core or optional context:
\[
\mathrm{CtxEff}
=
\frac{
|L(P)\cap (L(R_{\mathrm{core}})\cup L(R_{\mathrm{opt}}))|
}{
|L(P)|
}.
\]
Noise rate is the fraction of predicted regions that overlap neither core nor
optional context. The main table reports region-level noise.

\paragraph{Aggregation.}
Metrics are computed per instance and then averaged over instances. Empty
predictions receive zero for coverage, ranking, first-hit, and efficiency
metrics. Predictions with invalid paths or empty intervals are discarded before
scoring.

\section{Restricted-Context Validation Protocol}
\label{app:bridge}
\label{app:restricted_context}

This appendix supplements \S\ref{sec:bridge} and \S\ref{sec:downstream}. The
restricted-context protocol is used to test whether the selected regions can
support actual patch generation under a fixed patching scaffold; it is not part
of the standard upstream scoring loop.

\paragraph{Context materialization.}
For each explorer output, we first normalize paths, clip intervals to file
boundaries, and remove invalid regions. For selected files, only the predicted
line intervals remain visible. Lines outside the selected intervals are replaced
with blank placeholders rather than deleted, so that repository paths and line
numbers remain stable during patch generation and debugging. Files not selected
by the explorer are hidden from the patching agent.

\paragraph{Fixed patch scaffold.}
All restricted-context runs use the same patcher, prompt template, tool set, and
interaction budget. The only variable is the visible context produced by the
explorer. This control is important because otherwise a higher resolve rate
could reflect a stronger patch-generation scaffold rather than better
exploration.

\paragraph{Patch application and harness evaluation.}
After patch generation, the predicted diff is applied to the original repository
checkout and evaluated with the source benchmark's executable harness. An
instance is counted as resolved only when the patch passes the benchmark's
standard tests. Empty patches, unparsable diffs, failed patch applications, and
test failures are all counted as unresolved.

\paragraph{Failure diagnostics.}
For every unresolved run, we log a coarse failure reason: no diff generated,
invalid diff, patch failed to apply, patch outside visible context, applied patch
failed tests, timeout, or infrastructure error. These diagnostics do not change
the resolved/unresolved label; they are used to understand how restricted
context affects repair.

\section{Explorer Implementation Details}
\label{app:explorers}

This appendix supplements the explorer setup in \S\ref{sec:main-results}. All
methods are converted to the same output contract before scoring: an ordered
list of at most $K=5$ repository-relative line regions. Each region is represented
as a path and a closed line interval. Invalid paths, empty intervals, and regions
outside the repository checkout are discarded before evaluation.

\paragraph{Retrieval baselines.}
BM25 and TF--IDF use the issue statement as the query and rank repository chunks
by lexical similarity. The top-ranked chunks are converted into line regions.
Potion uses the same chunk-and-rank interface as a lightweight dense retrieval
baseline.

\paragraph{Agentic explorers.}
For agentic explorers, each system is run under its original search or
localization scaffold. We then normalize the resulting file, function, or region
outputs into line-level regions. When a method produces file-level outputs, we
map them to the most specific span supported by the method output or associated
read trace. This conversion is applied before scoring so that all methods are
compared under the same ranked-region contract.

\paragraph{Output validation.}
Before scoring, each prediction is checked for path validity, interval validity,
and repository membership. Invalid predictions are dropped. Predictions with
empty intervals are dropped. The remaining predictions are scored according to
their original order.

\section{Case Study: \texttt{scikit-learn/scikit-learn\#10844}}
\label{appendix:case-study}

To make the quantitative results in \S\ref{sec:main-results} concrete, we walk through a single instance for which all 14 explorers produced output and whose ground truth is small enough to inspect end-to-end. We pick a real numerical-overflow bug whose ranking on this single instance closely tracks the ordering in the main tables, so the case mirrors --- rather than distorts --- the global picture.

\paragraph{Issue (\texttt{scikit-learn/scikit-learn\#10844}, from SWE-Bench Verified).}
The issue is titled \emph{``\texttt{fowlkes\_mallows\_score} returns \texttt{RuntimeWarning} when variables get too big.''} The reporter observes that the line
\begin{center}\small\texttt{return tk / np.sqrt(pk * qk) if tk != 0. else 0.}\end{center}
inside \texttt{sklearn/metrics/cluster/supervised.py} silently overflows when \texttt{pk * qk} exceeds $2^{32}$, because \texttt{pk} and \texttt{qk} are 32-bit integers; the result is a \texttt{RuntimeWarning} and a corrupted score. The fix casts the operands to \texttt{np.float64} before the multiplication, plus adds a regression test in \texttt{tests/test\_supervised.py} that triggers the overflow path.

\paragraph{Ground truth.}
The refined ground truth lists two core files spanning only $26$ lines in total, so the line-recall view is not dominated by wide whole-file scopes:
\begin{center}\small
\begin{tabular}{lll}
\toprule
Path & Line range & Role \\
\midrule
\texttt{sklearn/metrics/cluster/supervised.py}              & 850--870  & modified function (\texttt{fowlkes\_mallows\_score}) \\
\texttt{sklearn/metrics/cluster/tests/test\_supervised.py}  & 245--249  & 5-line regression test for the overflow case \\
\bottomrule
\end{tabular}
\end{center}

\paragraph{Per-explorer outputs.}
Table~\ref{tab:case-sklearn} reports every explorer's top-5 output and the resulting metrics. Regions are abridged to fit the column; entries in [\,brackets\,] mark a region whose file overlaps a ground-truth file.

\begin{table}[h]
\centering
\small
\setlength{\tabcolsep}{4pt}
\caption{Top-5 outputs and metrics on \texttt{scikit-learn/scikit-learn\#10844}.
HF = HitFile, Noise = NoiseFile, Rec$_\ell$ = line recall, F1 = line F1, Cov = weighted core coverage.}
\label{tab:case-sklearn}
\begin{tabular}{l p{6.6cm} ccccc}
\toprule
Explorer & Returned regions (top-5, abridged) & HF$\uparrow$ & Noise$\downarrow$ & Rec$_\ell\uparrow$ & F1$\uparrow$ & Cov$\uparrow$ \\
\midrule
Oracle         & [supervised.py:850--870], [test\_supervised.py:245--249]                                                                       & 1.00 & 0.00 & 1.00 & 1.00 & 1.00 \\
\midrule
Random         & unrelated files from other repos                                                                                              & 0.00 & 1.00 & 0.00 & 0.00 & 0.00 \\
TF--IDF        & ISSUE\_TEMPLATE.md, CONTRIBUTING.md, doc/support.rst, doc/faq.rst                                                              & 0.00 & 1.00 & 0.00 & 0.00 & 0.00 \\
Potion (RAG)   & ISSUE\_TEMPLATE.md, CONTRIBUTING.md ($\times$2), doc/support.rst, sklearn/\_\_init\_\_.py                                       & 0.00 & 1.00 & 0.00 & 0.00 & 0.00 \\
BM25           & ISSUE\_TEMPLATE.md, PR\_TEMPLATE.md, CONTRIBUTING.md ($\times$2), [supervised.py:781--860]                                     & 0.50 & 0.75 & 0.42 & 0.07 & 0.31 \\
\midrule
AutoCodeRover  & [supervised.py:787--859]  (only 1 region emitted)                                                                              & 0.50 & 0.00 & 0.38 & 0.20 & 0.29 \\
OrcaLoca       & [supervised.py:787--859]  (only 1 region emitted)                                                                              & 0.50 & 0.00 & 0.38 & 0.20 & 0.29 \\
LocAgent       & [supervised.py:787--859], [supervised.py:53--107], [supervised.py:34--50]                                                      & 0.50 & 0.00 & 0.38 & 0.12 & 0.29 \\
CoSIL          & [supervised.py:1--$\infty$], cluster/setup.py, metrics/setup.py, cluster/\_\_init\_\_.py, metrics/\_\_init\_\_.py             & 0.50 & 0.80 & 0.81 & 0.05 & 0.60 \\
Claude Code    & [supervised.py:28--31/53--107/787--859/193--214], [test\_supervised.py:239--276]                                               & 1.00 & 0.00 & 0.58 & 0.14 & 0.69 \\
Mini-SWE-Agent & [supervised.py:787--859/53--95], [test\_supervised.py:239--276], doc/clustering.rst, doc/whats\_new/v0.18.rst                  & 1.00 & 0.50 & 0.58 & 0.14 & 0.69 \\
AweAgent       & [supervised.py:28--31/53--107/787--859], [test\_supervised.py:239--276], doc/clustering.rst                                    & 1.00 & 0.33 & 0.58 & 0.15 & 0.69 \\
Codex          & [supervised.py:852--859/53--107/579--605], [test\_supervised.py:239--276/170--184]                                             & 1.00 & 0.00 & 0.50 & 0.15 & 0.63 \\
OpenHands      & [supervised.py:787--859/53--107], [test\_supervised.py:239--276/170--184], metrics/classification.py                            & 1.00 & 0.33 & 0.58 & 0.14 & 0.69 \\
\bottomrule
\end{tabular}
\end{table}

\paragraph{Discussion.}
Three patterns are visible, and all three echo the global ordering reported in \S\ref{sec:main-results}.

\emph{First, one-shot lexical retrieval is essentially useless on this bug.}
TF--IDF, Potion, and Random all reach HitFile $=0$. The issue text contains no rare identifier that pins down the modified module: it talks about \texttt{RuntimeWarning}, overflow, and integer multiplication, all of which are far more frequent in the project's templates and documentation than in the implementation. BM25 partially recovers (HitFile $=0.50$) by anchoring on \texttt{fowlkes\_mallows\_score} appearing both in the title and in the file, but it still buries the function below four template files and never reaches the regression test. This is the regime the global numbers describe: sparse retrievers cluster near Random on file hit, and Potion adds little above them.

\emph{Second, the academic localizers behave like the rest of the paper: precise on the implementation file, blind to the test file.}
AutoCodeRover, OrcaLoca, LocAgent, and CoSIL all reach \texttt{supervised.py} and land squarely on the \texttt{fowlkes\_mallows\_score} block, but none of them surfaces \texttt{test\_supervised.py}; their HitFile therefore caps at $0.50$. AutoCodeRover and OrcaLoca additionally under-fill the top-5 budget, emitting only one region, which costs them line recall and coverage even within the file they do reach. CoSIL again emits whole files, which lifts Rec$_\ell$ to $0.81$ at the cost of high file-level noise --- the same Rec$_\ell$/HitFile asymmetry it shows in the main table.

\emph{Third, all five general-purpose agents converge on the same fix neighborhood.}
Claude Code, Mini-SWE-Agent, AweAgent, Codex, and OpenHands all reach both ground-truth files (HitFile $=1.00$), and all land on the \texttt{fowlkes\_mallows\_score} body in \texttt{supervised.py:787--859} plus the surrounding test block in \texttt{test\_supervised.py:239--276}. Differences between them reduce to small variations in span shape and how many auxiliary regions they include: Codex emits the tightest spans on the fix line itself (\texttt{852--859}), while AweAgent and OpenHands include extra context such as imports or related metrics. This mirrors the main-table observation that the general agents form a tight cluster on HitFile and Cov, with the remaining variance dominated by how much surrounding code each one chooses to return.

\emph{Takeaway.}
On this instance the practical ranking is Random/TF--IDF/Potion $\ll$ BM25 $<$ academic localizers (one file only, precise) $<$ general agents (both files, function-scoped spans) $<$ CoSIL (recall by whole-file emission) $<$ Oracle. The relative ordering and the magnitude of the gaps both match the global tables: the lexical baselines sit at the floor, the academic localizers occupy a narrow precision-leaning band, the general agents share a higher operating point on file hit and coverage, and only Oracle achieves matching scores on \emph{both} line-level and file-level metrics. The case therefore illustrates why we report file hit, line recall, and coverage together: each metric independently reproduces a different part of this same ordering.

\section{Reproducibility, Compute, and Limitations}
\label{app:reproducibility}

This appendix collects implementation information that is not central to the
main argument but is needed to interpret the released artifact and compute cost.

\paragraph{Reproducibility.}
The released artifact contains benchmark records, the common explorer-output
schema, metric computation scripts, and restricted-context validation scripts.
Each instance records source provenance, repository metadata, and line-level
context annotations. The evaluation pipeline consumes ranked-region predictions
and produces both per-instance metrics and aggregate tables.

\paragraph{Compute.}
Sparse retrieval baselines run on CPU workers after repository indexing. Dense
retrieval additionally requires embedding computation but no fine-tuning.
Agentic explorers and restricted-context validation are the most expensive
components because they require LLM calls and executable harness runs. For each
downstream run, we log the model, prompt, tool budget, wall-clock time, patch
status, and resolved status.

\paragraph{Limitations.}
SWE-Explore covers instances solved by at least one agent in our pool, so it does
not represent the full distribution of unsolved repository-level issues. Its
trajectory-derived ground truth is an empirical approximation of useful context,
not a proof that no other evidence could support a valid solution. Some valid
solution paths may rely on different evidence than the successful trajectories we
observe. The restricted-context protocol should therefore be read as a controlled
validation of exploration metrics, rather than as an absolute measure of
patch-generation ability.

\paragraph{Responsible release.}
SWE-Explore is derived from public software-engineering benchmarks and public
repository metadata. The release excludes private repositories, credentials, and
user data. Benchmark records preserve source attribution and include
documentation for schema, provenance, and intended use.

\end{document}